\documentclass[conference]{IEEEtran}
\usepackage{blindtext, graphicx}
\usepackage{graphics}
\usepackage{comment}
\usepackage{color}
\usepackage{memhfixc}
\usepackage[some,center] {background}
\usepackage{a4wide}
\usepackage{indentfirst}
\usepackage{fancybox}
\usepackage{listing}
\usepackage{authblk}
\usepackage{array}
\usepackage{amsmath,lipsum}

\ifCLASSOPTIONcompsoc
    \usepackage[caption=false, font=normalsize, labelfont=sf, textfont=sf]{subfig}
\else

\usepackage[caption=false, font=footnotesize]{subfig}
\fi

\ifCLASSINFOpdf
  
\else
  
\fi

\title{A Simple Probabilistic Model With Extended Kalman Filter To Predict Multi-leak In Pipelines}
\author[1]{Radhika P}
\author[2]{Anu Mol Joy}

\affil[1]{MGMCEPS,Valanchery,India,radhikap1993gmail.com}
\affil[2]{Lambton College, Canada, anumoljoy94@gmail.com}

\begin{document}

\maketitle

\begin{abstract} 
Pipelines for water supply are susceptible to burst-leakage due to fluid pressures of various nature. High pressure heads resulting in circumferential and (or) axial stresses larger than the material yield stress could cause pipe failure. Of equal concern is the local boiling or cavitation effect in regions of fluid pressure dropping below its vapor pressure, which in turn develop air bubbles that get transported through the pipeline, bursting later at remote locations. We initially developed a simple probabilistic model based on Method of Characteristics (MOC) to simulate burst leakage in pipelines, and compared with a pure deterministic hydraulic model. We had not considered cavitation effects for simplicity. The results indicated that the simple probabilistic model was only marginally different in its prediction of the transients on comparison with the latter. In order to determine the position and amount of leakage in the distribution system, the detection method based on simulating hydraulic transients was further evaluated using Extended Kalman Filter (EKF). We found that this non-linear filtering approach on the fluid transient model considerably reduced the number of input parameters required, and it was able to predict leakage rate and burst positions even in a highly noisy environment.\\\\
 \textit{Keywords\textemdash  Method of characteristics, Signal processing, CFD, Pipe burst, Numerical simulation, Probability}

\end{abstract}

\IEEEpeerreviewmaketitle
\section{INTRODUCTION}

Water distribution systems fail to achieve effective conservation and operational efficiency mainly due to leakages- small and large that goes undetected. Such leakages can result in serious economic losses as well as wastage of a very precious resource. Origin of the fluid transients in a pipe flow that cause burst leakage are mostly associated with the compressibility effects from sudden closure of valves (water hammer), sharp bends, sudden contraction (and) or expansion, etc. Hence, detection or early prediction of burst locations and leakage rates could be highly beneficial. 
\par Till date, different techniques have been used for the leak detection in water distribution systems. In 1985, Hargesheimer \cite{1}, used trihalomethanes (THMs) chloroform and dichlororbromomethane  in pipelines to find positions of water leakage. According to him, THM analysis provided a specific and sensitive means of identifying treated city water samples in seepage. However, this method works only on treated water and gives a general location of the leak but not its magnitude. 
\par In 1987, Dalle and Himmelblau \cite{2}  used the Kalman Filter for fault detection within a single stage evaporator.  Li and Olson [3], in 1991 applied the Extended Kalman Filter to a closed loop non-linear distillation process.
\par Liou and Tian \cite{4}, in 1995 developed a model for a single pipeline using transient flow simulations. They considered data noise in pressure and flow measurements and found that noise limits leak detectability.
\par In 1999, Brunone \cite{5} proposed a technique for leak detection in outfall pipes based on properties of transient pressure waves. The occurrence of transient damping determined the presence of a leak and the timing of the damping determined the location.
\par Vtkovsky et al \cite{6} in 2000, detected leaks in water distribution systems using the genetic algorithm (GA) technique in conjunction with the inverse transient method (ITM). The slow rate of convergence within complex systems is a disadvantage of this method.
\par In 2001, Mpesha et al \cite{7} tried a leak detection frequency response method, which required measuring pressure and discharge at one location in the pipeline as the input parameters.
 \par Buchberger \cite{8} in 2004, developed a statistical method for detecting the magnitude of leaks in pipe networks. Mean and standard deviation of the measured flows were computed, and the maximum network leakage rate was determined when the flow values diverge from the statistical curves. But this method does not locate the position of the leak.  Verde \cite{9} developed a method for location of leak in a pipeline, by using flow and pressure sensors only at the ends of the pipeline. According to him, it can be solved using a simple nonlinear model of the flow, assuming leak position with uncertainty, and combining static relationship between residual components and leak position error.
\par In 2005, Misiunas et al \cite{10} tried a method to find leaks in the pipeline by measuring pressure at one location to sense the negative pressure wave that was produced when a break occurred. The location was found by the timing of the initial and reflected transient waves produced by the break.
\par Lesyshen \cite{11} in 2005, used a model based on single, fictitious leak for leakage detection. The objective his method was to determine through simulations, the effectiveness of the Extended Kalman Filter for such problems.
\par Doney \cite{12} in 2007, also used a model based algorithm to detect a leak in a pipeline. The model was able to detect the location and magnitude of a leak in a pipeline accurately once four pressure measurements were inputted into the EKF.
 \par Cataldo and Cannazza \cite{13} in 2012, presented a time domain reflectometry (TDR) -based system for leak detection in underground metal pipes, which considerably reduced time required for inspection.
 \par In 2015, Golmohamadi \cite{14} applied both hardware-based and software-based techniques for leak detection in pipelines. Emission of ultrasonic wave was used for pipeline inspection, while his software method was based on the hydraulic transient model for the pipeline. He concluded that the hardware based method was reliable but very expensive in leak detection and was applicable only in shorter ranges compared to the software based approach.
 \par In 2015, Aguinaga et al \cite{15} proposed a model based approach to detect and isolate non-concurrent multiple leaks in a pipeline using an Extended Kalman Filter as a state observer. But this approach was valid only if the number of leaks were not very large.
 \par In 2018, Khalilabad et al \cite{16} developed a technique based on hydraulic model to determine leakage in distribution system using Extended Kalman Filter (EKF). The results showed that the EKF could determine amount and position of leakage on pipeline with significantly less error.
\par The prime objective of our method is to improve the Lesyshen’s single fictitious leak detection model \cite{11} into a probabilistic multiple leak detection method based on Hoop stress (HS) and yield stress (YS) developed in the pipelines. Considering that high pressure heads resulting in circumferential (hoop) stress or longitudinal (axial) stress will be comparable to material yield stress during a burst-leakage, we have based our model on probability to allow only a few among such eligible locations to burst. Again, we have neglected the effect of cavitation for simplicity. Further, this hydraulic transient based model was evaluated using an EKF scheme to give the state space estimates in a highly noisy environment. This is required because, in the event of multiple leaks, the pipeline system would generate very noisy pressure signals for over a wider range. However, the EKF technique requires adequate monitoring of only a few selected pressure heads as inputs into a state space estimation scheme, to generate meaningful results.
\par The remainder of this paper is organized as follows. The two different cases to model, a horizontal and a non-horizontal pipeline with differing boundary conditions, are presented in section II. Section III presents the theory behind transient fluid flow simulation for pipelines. Detailed methodology is discussed in Section IV. The theory of Extended Kalman Filter, which combines the transient fluid flow theory of Section III and the EKF technique for the current model, is given in Section V. The results are presented in Section VI.

\section{}
\subsection{THE CASE OF HORIZONTAL PIPELINE}

\begin{figure}[h]
  	  \centering
  		\includegraphics[width=0.5\textwidth]{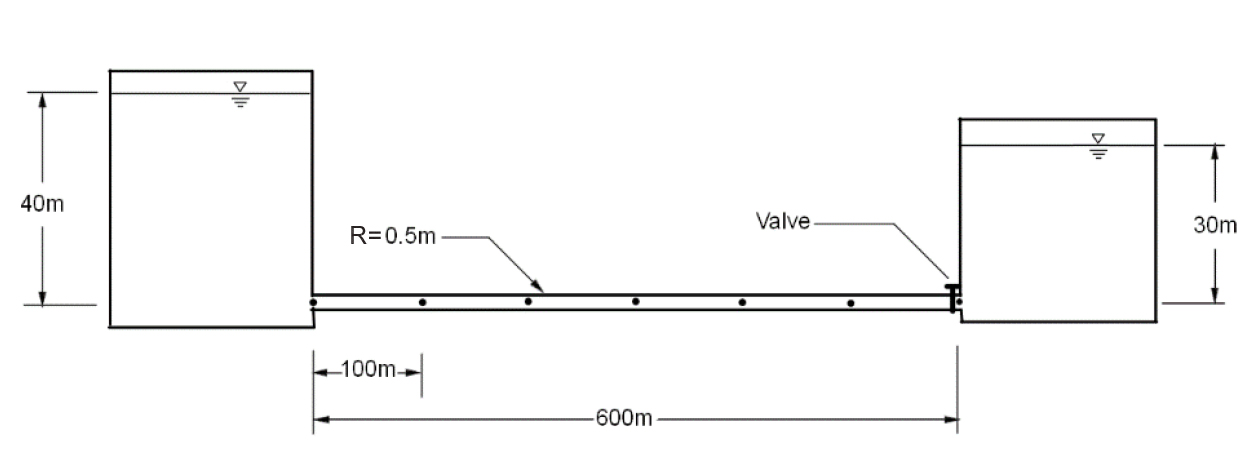}
  	  \caption{ The case of horizontal pipeline layout}
  	  \label{figure:f2}
      \end{figure}

The case of horizontal pipeline consists of a constant head reservoir maintained at 40m head connected to a downstream reservoir maintained at 30m through horizontally laid pipelines 600m long and which are 0.5m in diameter. A gate valve is located just upstream of the 30m reservoir to regulate the flow. The pipeline is segmented into six equal pipe sections which are 100m long each. The end location of a pipeline is referred to as a ‘node’. Hence there are seven nodes: one upstream at supply reservoir, five interior nodes equally spaced at 100m from the supply, and one end node at the valve just before the downstream reservoir. This configuration according to the work of Leyshan [11] is as shown in Figure \ref{figure:f2}.

\subsubsection{Model parameters for II.A}

The parameters of each node in the horizontal pipeline consist of elevation, pressure head and flow rate. The initial variables to the model are as given in Table \ref{table:1}. The gate valve located just ahead of the lowest reservoir is allowed to close in 20 seconds.   

\begin{table}[h!]
\centering
\begin{tabular}{ |c|c| } 
 \hline
Density $\ (kg/m^3) $\ of the fluid & rho=1000	\\
\hline
Acceleration due to gravity $\ (m/s^2) $\ & g=9.8 \\
\hline
Fluid modulus of elastisity $\ (Pa) $\  & $\ \beta $\ =2.1994e9 \\
\hline
Length of each pipe $\ (m) $\ & l=100 \\
\hline
Diameter $\ (m) $\ & d=0.5 \\
\hline
Friction factor & f=0.015 \\
\hline
Pipe wall thickness $\ (m) $\ & e=0.01905 \\
\hline
Youngs modulus of elasticity of pipe & E=4.1e11 \\
\hline
Yield stress $\ (N/m^2) $\ & Y=8e6\\
\hline
Upstream Head $\ (m) $\   & $\ H_{R1} $\ =40 \\
\hline
Downstream Head $\ (m) $\   & $\ H_{R2} $\ =30 \\
\hline  
Valve closure time $\ (s) $\ & $\ t_c $\ = 20    \\
\hline
  
\end{tabular}

\caption{Input data for horizontal layout}

\label{table:1}
\end{table}

\subsubsection{Origin of transients for II.A}

Initially, the regulating gate valve is fully open while allowing flow from the upstream reservoir to the downstream reservoir. This valve is closed suddenly in steps which cause pressure surge ('water hammer') in the pipeline. Water hammer due to sudden deceleration of fluid flow can generate very high pressure transients which could burst a pipeline and can generate pipeline vibrations.
\par Hence the steady state condition for this case is defined by the fully developed pipe flow with downstream gate valve opened to its full position. On closure of valve in steps, the variation from the steady state values for the mutually related unknowns, viz., flow rate in each pipe, \mbox{Q [$\ m^3/s $\ ]} and the total energy head at each junction node, H [$\ m $\ ] would mathematically describe the flow and pressure distribution within the pipe under unsteady state conditions. The fundamental relationship between conservation of mass and energy for the hydraulic transmission will define the same.

\subsection{THE CASE OF NON-HORIZONTAL PIPELINE}

Another fictitious case with a non-horizontal layout and different boundary conditions is as follows. This distribution setup consists of five pipes each 20m in length and 0.5m diameter. A reservoir with a constant head of 60m is present upstream to the first pipe. The end of pipeline is closed allowing no flow past and hence the fluid is stagnant. There are 6 nodes in this configuration: a node at upstream supply reservoir, four interior nodes and last being at the downstream closed end. 
\subsubsection{Model parameters for II.B}
\begin{figure}[h]
  	  \centering
  		\includegraphics[width=0.5\textwidth]{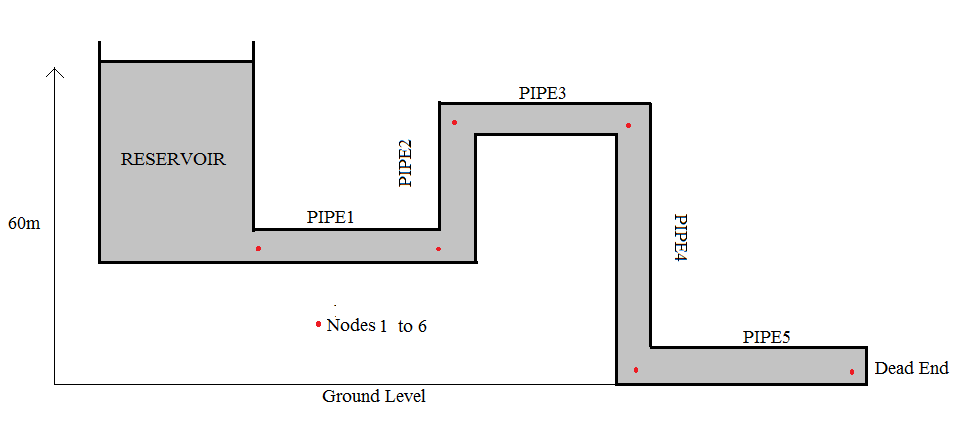}
  	  \caption{ The case of horizontal pipeline layout}
  	  \label{figure:fnon}
      \end{figure}

\begin{table}[h!]
\centering
\begin{tabular}{ |c|c| } 
 \hline
Node ID&  Elevation (m)\\
\hline
1&  	20\\  \hline
2&      20\\   \hline 
3&  	30\\   \hline
4&  	30\\	 \hline  
5&  	0\\	   \hline
6&      0\\    

\hline 
\end{tabular}
\caption{The node elevations}
\label{table:elevation}
\end{table}

\begin{table}[h!]
\centering
\begin{tabular}{ |c|c| } 
 \hline
Density  [$\ kg/m^3 $\ ] of the fluid &	rho=1000 \\
\hline
Acceleration due to gravity [$\ m/s^2 $\ ]   &	g=9.811	 \\ 
\hline
Length of each pipe $\ (m) $\ & l=100 \\
\hline
Diameter $\ (m) $\ & d=0.5 \\
\hline
Fluid modulus of elastisity [$\ Pa $\ ]& $\ \beta $\ =$\ 2.1994e^9 $\            \\ \hline
Friction factor& f=0.015           \\ \hline
pipe wall thickness [m]& e=	0.01905            \\ \hline
Youngs modulus of elasticity of pipe & E= 4.1e11           \\ \hline
Yield Stress & Y= $\ 8e^6$\ 
    \\ \hline
Upstream head [m] & $\ H_{R1} $\ = 40
        \\ \hline
\end{tabular}
\caption{Input data for non-horizontal layout}
\label{table:input}
\end{table}

\subsubsection{Origin of transients for II.B}

Through the case of non-horizontal layout we tried to understand the robustness of our model in handling a completely different set of boundary conditions and fluid transients.  In this problem, we allowed for the burst leakage of a fictitious pipe material which had it its pressure head solely due to the supply reservoir. The fluid was stagnant initially in the entire pipeline, and the fluid transients were supposed to be triggered by the onset of burst leakage.
\par Hence contrary to the former case, the steady state conditions for this case were initially calculated based on static pressure head, and the transients Q and H, were modeled on the onset of burst leakage.

\section{GOVERNING EQUATIONS FOR TRANSIENT FLOW}

Transient flow in pipelines can be explained by the equations for conservation of mass (continuity equation) and momentum (Navier-Stokes equation). Dissipation of frictional energy in the form of heat is not modeled in the present study. These equations are represented by partial differential equations in time ($\ t $\ ) and space ($\ x $\ ). $\ H $\ and $\ Q $\ are considered as the dependent variables while t and x are independent variables. The goal is to determine the dependent variables as a function of time and space. 

\subsection{Continuity Equation}
The continuity equation is derived from the principle of mass conservation which states that for a control volume, the 'mass flow in' is equal to 'mass flow out' from it. For fluid flow through an elastic pipe this equation takes the form \cite{17},

   \begin{equation}
      \frac{a^2}{gA}\frac{\partial Q}{\partial x}+\frac{\partial H}{\partial t}=0
      \end{equation}
      
 where, $\ A $\ is the pipe cross-sectional area [$\ m^2 $\ ], and $\ a $\ is the wave speed within the fluid medium inside the pipe and it is given by

 \begin{equation}
       a=\sqrt{\frac{\left (\frac{\beta}{\rho}\right )}{1 +\left (\frac{\beta}{E}\right) \left (\frac{D}{e}\right)c_1}}
       \end{equation}
 
 Here $\ c_1 $\ is a constant assuming pipe anchored with expansion joints throughout and
 $\  e   $\ is the thickness of the pipe walls [$\ m $\ ]. \\

\subsection{Transient Momentum Equation (Navier-Stokes Equation)}

The momentum equation stating the dynamical equilibrium of the fluid for incompressible fluid flow is given by
\begin{equation}
\frac{\partial Q}{\partial t}+gA\frac{\partial H}{\partial x}+\frac{f}{2DA}Q|Q|=0
\end{equation}
\\

\section{METHODOLOGY} 

In order to solve the two hyperbolic governing equations for transient fluid flow, we have used the Method of Characteristics (MOC); a robust method used extensively for solving partial differential equations in engineering simulation and is found to provide accurate results. Through MOC, the partial differential equations were transformed into two ordinary differential equations. A simple probabilistic method was coupled to this pure hydraulic model to determine the position and amount of leakage in distribution system. This detection method was further evaluated using the non-linear Extended Kalman Filter (EKF) under a very noisy environment.

\par For more details on MOC and boundary conditions for hydraulic transients discussed in the following sections, the reader is advised to refer Chaudhary []. A short description for the individual elements of the numerical scheme can be as follows. 

\subsection{The method of characteristics}

The method of characteristics converts the two partial differential equations of momentum and continuity into four ordinary differential equations.

\begin{equation}
\frac{dQ}{dt}+\frac{gA}{a}\frac{dH}{dt}+\frac{f}{2DA}Q|Q|=0
\label{equation:a8}
\end{equation}\\
if 
\begin{equation}
\frac{dx}{dt}=a
\label{equation:a9}
\end{equation}
and 
\begin{equation}
\frac{dQ}{dt}-\frac{gA}{a}\frac{dH}{dt}+\frac{f}{2DA}Q|Q|=0
\label{equation:a10}
\end{equation}
if\\
\begin{equation}
\frac{dx}{dt}=-a
\label{equation:a11}
\end{equation}
 
The compatibility equations \ref{equation:a8} and \ref{equation:a10} (C+ and C- equations) exist along C+  and  C- characteristic lines respectively. The physical significance of characteristics lines is that, if the pressure head and flow at point A are known, then equation \ref{equation:a8} can be integrated from point A to point P. Then the resulting equation will be in terms of unknown pressure head and flow at point P. Similarly knowing the pressure head and flow at point B, equation \ref{equation:a10} can be integrated along line BP which results in a equation relating pressure head and flow at point P. Thus at point P, there are two equations and two unknowns (pressure head and flow) which can be solved. In this way pressure head and flow at each point can be calculated throughout time.

\subsection{Discretization}
\normalsize Numerical discretization allows for solution of governing partial differential equations based on discrete computational points (nodes). Integrating equation \ref{equation:a8} along the $\ C+ $\ characteristic line (AP) and solving for flow at point P gives,

\begin{equation}
    Q_P=C_p-C_aH_P
    \label{equation:a18}
    \end{equation}
    
where  \begin{equation}
        C_p=Q_A+\frac{gA}{a}H_A-\frac{f\Delta t}{2DA}Q_A|Q_A|
        \label{equation:a20}
        \end{equation}

Similarly, integrating equation \ref{equation:a10} along the $\ C- $\ characteristic line (BP) and solving for flow at point P gives,
 
 \begin{equation}
    Q_P=C_n+C_aH_P
    \label{equation:a19}
    \end{equation}

where,

\begin{equation}
    C_n=Q_A-\frac{gA}{a}H_B-\frac{f\Delta t}{2DA}Q_B|Q_B|
    \label{equation:a21}
    \end{equation}
    
and 
 
 \begin{equation}
        C_a=\frac{gA}{a}
        \label{equation:a22}
        \end{equation}
        
The values of two unknowns ($\ Q_P $\ and $\ H_P $\ ) can be determined by simultaneosly solving  equations \ref{equation:a18} and \ref{equation:a19}, i.e.,

            \begin{equation}
            Q_P=\frac{C_p+C_n}{2}
            \label{equation:a23}
            \end{equation}
            
The value of $\ H_P $\ can be determined by using either equation ~\ref{equation:a18} or ~\ref{equation:a19}.  Thus, by using equation ~\ref{equation:a18} and  ~\ref{equation:a19}, conditions at all interior points at the end of each time step can be determined. However, at the boundaries, either equation ~\ref{equation:a18} or ~\ref{equation:a19} is availabe. So, special conditions are needed to determine the parameters at the boundaries at time $\ t_0+\Delta t $\ \ \cite{17} .

 \subsection{Boundary Conditions}
 The boundary condition for the upstream reservoir
  is obtained from the energy equation. It has to be noted that, the boundary condition for the downstream reservoir is obtained from the valve equation for case II.A and from dead end condition for case II.B. The special boundary conditions which need to be taken care are as discussed below.
  
 \subsubsection{The Supply Reservoir}
 For the upstream reservoir at the beginning of water distribution line (node 1), the boundary condition is given as
 
 \begin{equation}
  H_{1,k}=H_{R1}-(1+\eta)\frac{Q_{11,k}^2}{2gA^2}
  \end{equation}
  
  \begin{equation}
  Q_{11,k}=\frac{-1+\sqrt{1+4K_1(C_n+C_aH_{R1})}}{2K_1}
  \label{equation:a28}
  \end{equation}
  
  where,
  
  \begin{equation}
  K_1=\frac{C_a(1+\eta)}{2gA^2}
  \label{equation:a29}
  \end{equation}
  
  Here$\ H_{R1} $\ represents the head at the supply reservoir and $\ \eta $\  is the entrance loss coefficient, taken as 0.5.
  
 \subsubsection{The Downstream Reservoir with Valve}
 
  There is a valve at the downstream reservoir for first case (II.A) and the boundary condition for the downstream reservoir is modeled using the valve equation. The orifice equation for steady state flow through a valve is \ \cite{17}:
  
  \begin{equation}
  Q_{62,o}=(C_{do}A_{vo})\sqrt{2g(H_{6,o}- H_{R2})}
  \label{equation:a35}
  \end{equation}
  
 where $\ Q_{62,o} $\ is the steady state flow through the valve, $\ (H_{6,o}- H_{R2}) $\  is the steady state
  head loss across the valve, $\ H_{R2} $\ is the downstream reservoir head, $\ C_{do} $\  is the steady state
  discharge coefficient and here it is chosen as 0.6 and $\ A_{vo} $\  is the pipe area when valve is opened fully. For a general
  opening the flow may be described as:
  
  \begin{equation}
  Q_{62,k}=(C_dA_v)\sqrt{2g(H_{6,k}-H_{R2})}
  \label{equation:a36}
  \end{equation}
  
  Here $\ A_{v} $\ is the instantaneous valve opening area and $\ C_{d} $\ is the coefficient of discharge in transient condition \cite{15}, that is used to characterise the flow and pressure head behaviour at the orifices and it is given by \ \cite{19},
  
  \begin{equation}
  C_d=0.05959+0.0312 \alpha ^{2.1}-0.184 \alpha ^6
  \label{equation:aa36}
  \end{equation}
  
 where, $\ \alpha $\ is the area ratio ($\ D_0^2/D_1^2 $\ ), $\ D_1 $\ is the diameter of the pipe when valve is opened fully and  $\ D_0 $\ is the constricted diameter.\\ 
 
 \par Solving for flow at the valve ( $\ Q_{62,k} $\ ) gives \ \cite{17},
 
 \begin{equation}
 Q_{62,k}=\frac{-C_{v}+\sqrt{C_{v}^2-4 C_{v} (C_{a}H_{R2}-C_{p})}}{2}
 \end{equation}

 in which
 
 \begin{equation}
 C_v=\frac{(\tau Q_o)^2}{C_a(H_{6,o}- H_{R2})}
 \label{equation:a40}
 \end{equation}
 
 Here $\ Q_o $\ is the steady state flow through the valve and $\ \tau $\ is the dimensionless valve opening area and given by 
 
 \begin{equation}
 \tau = \frac{(C_dA_v)}{(C_{do}A_{vo})}
 \label{equation:a38}
 \end{equation}
 
\begin{figure}[h]
      	  \centering
      		\includegraphics[width=0.5\textwidth]{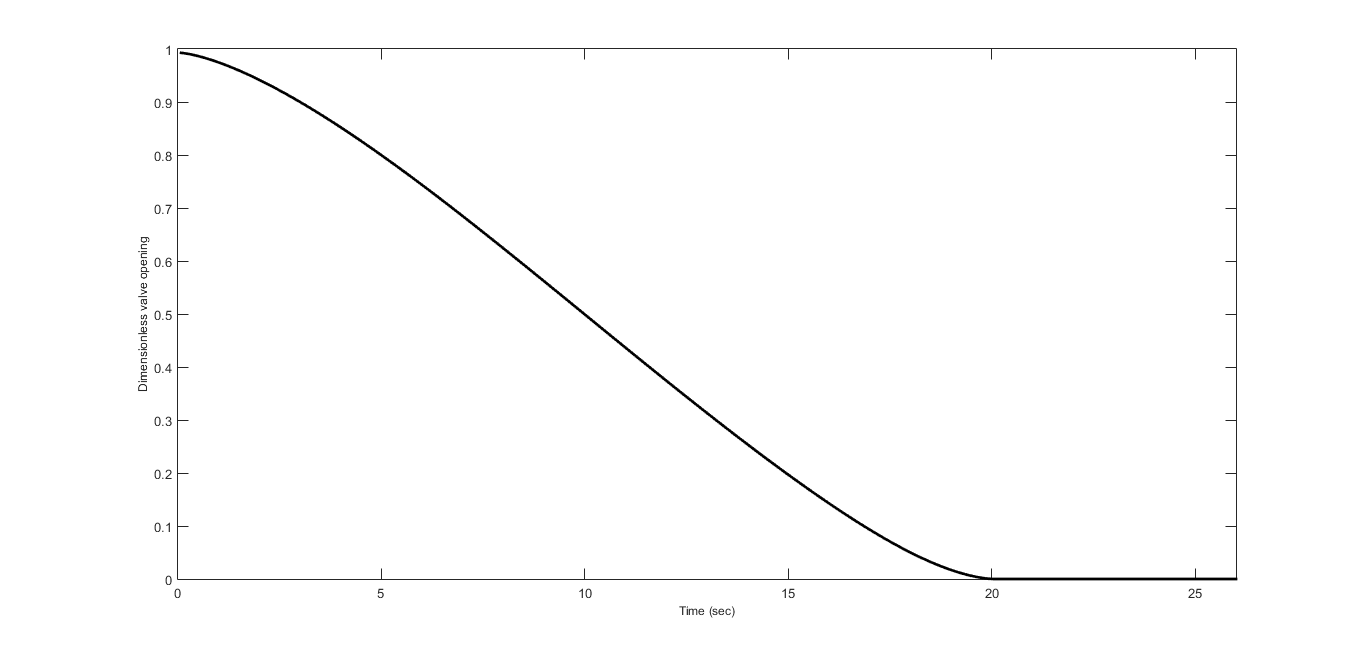}
      	  \caption{Variation of non-dimensionalized valve closure time }
      	  \label{figure:f4}
          \end{figure}

 For a fully open valve, $\ \tau $\ =1. The variation of $\ \tau $\ over closure time $\ t_c $\ can be as given in Figure \ref{figure:f4} 
  
\subsubsection{The downstream closed pipe end}          
          
The pipeline in the second case (II.B) is blocked at downstream end. This dead end boundary condition gives, flow rate as zero, i.e.,
\begin{equation}
  Q_p  =0
  \end{equation}
  
  Hence, from the positive characteristic equation,
    \begin{equation}
     H_P=\frac{C_p}{C_a}
     \end{equation}

\subsubsection{Inner Nodes with Leakage}  

\begin{figure}[h]
	  \centering
		\includegraphics[width=0.5\textwidth]{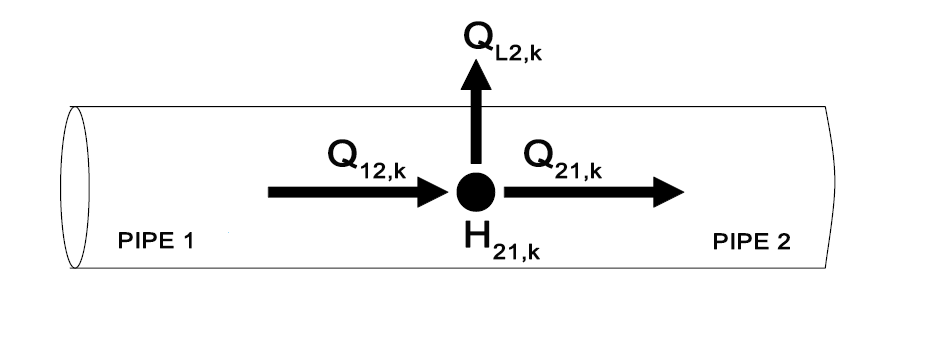}
		\caption{Flow configuration with leakage at node 5}
	\label{figure:flow}	
\end{figure}

Apart from the above boundary conditions, modeling the pipe breakage would require additional equations for flow rate ($\ Q $\ ), pressure heads ($\ H $\ ) and leakage rate ($\ Q_L $\ ) as discussed below. On the onset of leakage in a pipe, fluid transients in the entire pipeline changes altogether when compared to the no-leak situation. Leakage at any interior location (node) can be modeled using an orifice equation. This serves as an additional boundary condition, wherein continuity equation is enforced at the leaking node. Suppose a leakage is detected at node 5 (Figure \ref{figure:flow}), then the continuity equation and characteristic wave equation using MOC changes as,

\begin{equation}
Q_{42,k}=Q_{51,k}+Q{L,5}
\end{equation}

\begin{equation}
Q_{42,k}=C_p-C_aH_{5,k}
\end{equation}

\begin{equation}
Q_{51,k}=C_n+C_aH_{5,k}
\end{equation}

The determination of leak rate at node 5 ($\ Q_{L,5} $\ ) is as discussed in section IV.D. 1

\subsection{Modeling Pipe Breakage} 

Initially the steady state condition for pressure head $\ (H) $\ and flow rate $\ (Q) $\ were evaluated based on steady state equations considering the reservoir heads maintained and fluid friction along pipelines. Fluid transients (unsteady pressure head and flow rates) in a pipe flow were allowed to develop due to sudden closure of valve (water hammer), sharp bends, sudden contraction (and) or expansion, etc. The local burst-leakages themselves were supposed to generate further transient signals. Hence, this coupled complex phenomenon is highly non-linear in nature and is essentially a problem of fluid-structure interaction and signal processing. For simplicity we have neglected the effects of cavitation in creating transients.
\par In real scenario, the high local pressure heads resulting from fluid transients may not cause burst leakage at all positions defined by a pure deterministic model, since leak initiated at a position may cause pressure relief at other locations. So, at all time steps, the eligibility of a location for burst leakage has to be assigned probabilistically. Hence, while simulating fluid transients, we decided to compare the pure deterministic hydraulic model with a simple probabilistic hydraulic model. 

\subsubsection{Deterministic Model}
For the first case (II.A), the valve at the downstream end was allowed to be closed in steps, creating sudden pressure rise. We considered that if the circumferential stress or axial stress on the pipeline generated by this unsteady pressure rise was larger than 80 \% of the material yield stress (YS), then it could cause the burst leakage. For the second case (II.B), we allowed the supply head reservoir to be maintained at such a head that at least a few locations of the non-horizontally laid fictitious pipe material conform to the above burst criteria.
\par By this hypothesis, a set of plausible burst nodal locations were identified. In the deterministic model all such locations were allowed to break. The leak area $\ A_{leak} $\ was assumed to be non-variant (no creep allowed) for which a non-dimensionalized area parameter $\ \lambda $\ (assumed to be 0.001) was considered to satisfy the empirical relation for leak rate $\ (Q_L) $\ as,

\begin{equation}
Q_L=  \lambda\sqrt{H} 
\end{equation}
where,
\begin{equation}
\lambda=  A_{leak}\sqrt{2g} 
\end{equation}

\subsubsection{Probabilistic Model}
 The main disadvantage of the above deterministic model could be the over prediction of the plausible breakage points. Hence a probabilistic method to choose the breaking locations based on some distribution function (normal, Gaussian, etc) is more suitable for such prediction. However, it may require exhaustive experimental data or previously obtained data from trial runs to decide upon this function. Hence, for the present work in which our prime objective is to evaluate the hydraulic model with EKF, we employ a relatively simple probabilistic method. This simple model is based on merely a coin-tossing probability to assign a location (node), from among the group of potentially plausible nodes, its eligibility to break. Further, we also analyze how different the prediction of a deterministic model could be from a simple probabilistic model. Condition for leakage according to the simple probabilistic method is as in Figure 5.

\par Condition for leakage:

\begin{figure}[h]
	  \centering
		\includegraphics[width=0.5\textwidth]{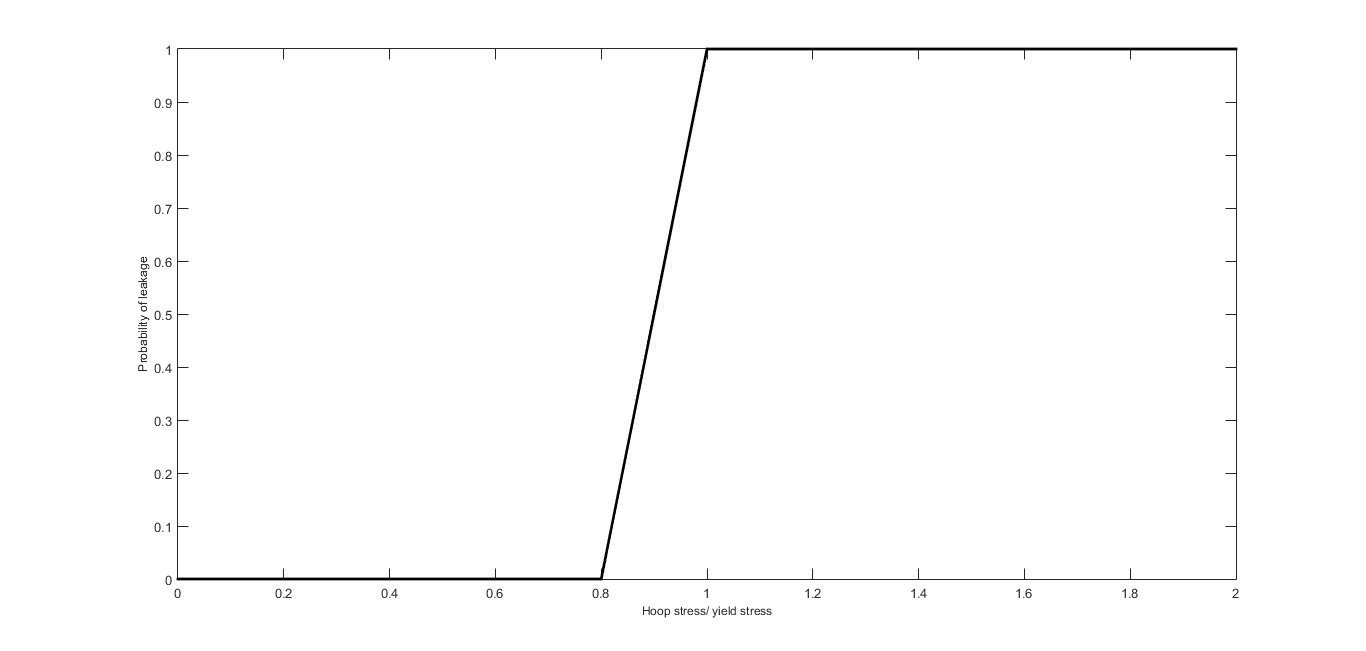}
	  
    \end{figure}
{\footnotesize 
\[\mbox{Probability\ (leakage)}= \left\{ \begin{array}{ll}
         0 & \mbox{if $HS < 0.8   YS$}\\
        1 & \mbox{if $HS >  YS$}\\
        \frac{(HS-0.8 YS)}{YS(1-0.8)} & \mbox{else}\end{array} \right. \]
}
\\

\subsection{Modeling Inner Nodes with Leakage}

Apart from the boundary conditions, modeling the pipe breakage would require additional equations for flow (or) and leakage rates and pressure heads as discussed below. When a leak is detected in a pipeline by any of the above methods, then the fluid transients in the
pipeline changes compared to the no-leak situation. The leakage at any interior node can be modeled using an orifice
equation. Continuity equation is enforced at this leaking node. Suppose a leakage is detected at node 5, then the transients in that pipeline with leaks have to be additionally solved numerically
using MOC. In such case the continuity equation and characteristic wave equation changes as,  \\ \bigskip

\begin{figure}[h]
	  \centering
		\includegraphics[width=0.5\textwidth]{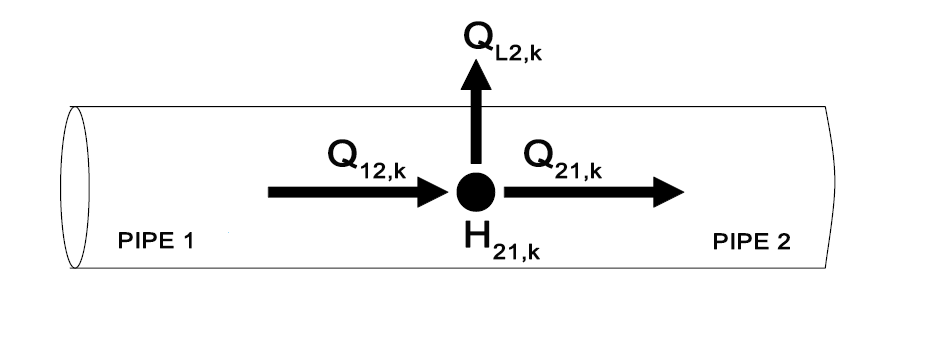}
		\caption{Leakage at Node 5}
\end{figure}

\begin{enumerate} 
\item Continuity Equation:

\begin{equation}
Q_{42,k}=Q_{51,k}+Q_{L5}
\end{equation}

\item Characteristic Equations:

\begin{equation}
Q_{42,k}=C_p-C_aH_{5,k}
\end{equation}

\begin{equation}
Q_{51,k}=C_n+C_aH_{5,k}
\end{equation}

\begin{equation}
Q_{L5}=\lambda_5\sqrt{H_{5,k}}
\end{equation}

\end{enumerate}
where,
\begin{equation}
C_p=Q_{32,k-1}+C_aH_{4,k-1}-\frac{f \Delta t}{2DA}Q_{32,k-1}|Q_{32,k-1}|
\end{equation}

\begin{equation}
C_n=Q_{52,k-1}-C_aH_{6,k-1}-\frac{f \Delta t}{2DA}Q_{52,k-1}|Q_{52,k-1}|
\end{equation}
 and $\ \lambda_5 $\ is the unknown leakage area constant for node 5 (chosen as .001).
\par Solving for Head at node 5 gives:
\begin{equation}
H_{5,k}=\left (\frac{\lambda_5^2}{8C_a^2} \right )+\frac{1}{2C_a}(C_p+C_n)-\frac{\lambda_5}{8C_a^2}\sqrt{\lambda_5^2+8C_a(C_p+C_n)}
\end{equation}

\section{Applying the Extended Kalman Filter for Leak Detection}

The Extended
Kalman Filter (EKF) is used to estimate pipeline leakage, given a pipeline
model and set of inputs (Head at node 1 and 7). Since the  model equations that represent the conditions in the pipeline
are non-linear, the EKF is implemented.

\subsection{The Filter Model and State Space Representation}

A state space representation of the describing equations is needed for filter implementation. The state vector can be described as:

\begin{align}
\mathbf{x}_k &= \begin{bmatrix}
           x_{1} \\
           x_{2}\\
          x_{3}\\
          x_{4}\\
          x_{5}\\
           x_{6}\\
           x_{7}\\
           x_{8}\\
           x_{9}\\
          x_{10}\\
          x_{11}\\
          x_{12}\\
          x_{13}\\
          x_{14}\\
          x_{15}\\
          x_{16}\\
          x_{17}\\
           x_{18}\\
           x_{19}\\
            x_{20}\\
          x_{21}\\
          x_{22}\\
          x_{23}\\
           x_{24}\\     
\end{bmatrix}=
\begin{bmatrix} 
	H_{1,k}\\
	H_{2,k}\\
	 H_{3,k}\\
	 H_{4,k}\\
	 H_{5,k}\\
	H_{6,k}\\
	H_{7,k}\\
	Q_{11,k}\\
	Q_{12,k}\\
	Q_{21,k}\\
    Q_{22,k}\\
    Q_{31,k}\\
    Q_{32,k}\\
    Q_{41,k}\\
    Q_{42,k}\\
    Q_{51,k}\\
    Q_{52,k}\\
	Q_{61,k}\\
	Q_{62,k}\\
	Q_{L2,k}\\
	Q_{L3,k}\\
	Q_{L4,k}\\
	Q_{L5,k}\\
	Q_{L6,k}\\  
\end{bmatrix} 
\end{align} 

The inputs into the model are the upstream and downstream head and the valve
coefficient.

\begin{align}
\mathbf{u}_k &= \begin{bmatrix}
        u_1\\
        u_2\\
        u_3\\      
\end{bmatrix}=
\begin{bmatrix}
	H_{R1}\\
	H_{R2}\\
	C_v
\end{bmatrix}
\end{align}

 The output equation is given by:

\begin{align}
\mathbf{z}_k &= \begin{bmatrix}
        x_{1,k}\\
        x_{7,k}\\      
\end{bmatrix}+v_k=
\setcounter{MaxMatrixCols}{24}
\begin{bmatrix}
1 & ...... & 0 & .... & 0\\
0 & ...... & 1 & .... & 0
\end{bmatrix}\mathbf{x}_k+v_k
\end{align} 

 where $\ v_k $\ represents measurement noise. The upstream and downstream pipeline heads ($\ H_{R1} $\  and $\  H_{R2} $\ ) are taken as inputs. The Head at node 1 and 7 are the measurements taken from the system.
\par In general, the non-linear stochastic difference equation, in state space form, is given as:

\begin{equation}
x_k=f(x_{k-1},u_{k-1})+w_{k-1}
\end{equation}

The implementation of EKF requires the equations to be linearized around the
current estimate and can be determined by computing the Jacobian matrix. The
Jacobian matrix is the rate of change within the state vector, with respect to each state. The Jacobian is given by equation \ref{equation:a60}

\begin{equation}
J_x=\frac{\partial f[x]}{\partial x}\lvert\hat{x}_{k}=
  \left[ {\begin{array}{ccccc}
   \frac{\partial f_1[x]}{\partial x_1}\lvert\hat{x}_{1} & \frac{\partial f_1[x]}{\partial x_2}\lvert\hat{x}_{2} & ... \\
    \frac{\partial f_2[x]}{\partial x_1}\lvert\hat{x}_{1} & \frac{\partial f_2[x]}{\partial x_2}\lvert\hat{x}_{2} & ... \\
    . & . & . \\
    . & . & & . \\
    . & . & & & . \\
  \end{array} } \right]_{\hat{x}_{k}}
  \label{equation:a60}
\end{equation}

 \subsection{Initial Conditions  and Covariance}
  The Initial conditions required for the Extended Kalman filter to start the estimation
 process are the measurement covariance matrices $\ Q_k $\ and $\ R_k $\ , the initial state estimates $\ \hat{x}_{0}^{-}  $\, and the a priori error covariance $\ P_{0}^{-} $\ . The initial state
 estimates were determined from a steady state analysis, assuming zero leakage in the water distribution line. The initial value of the a priori error covariance $\ P_{0}^{-} $\ is choosen as, $\ P_{0}^{-} = IC  $\ , where $\ I $\ is
 24x24 identity matrix and $\ C $\ a constant and here it is 0.1. For this work the error covariance matrices were set to:
 
 \[
 R_k=
   \begin{bmatrix}
     0.001 & 0 \\
     0 &  0.001
   \end{bmatrix}
 \]
 
 and,
 
 \[
 Q_k=
   \begin{bmatrix}
     0.1 \:I_7 & 0 & 0 \\
     0 &  0.01 \:I_{12} & 0 \\
     0 & 0 & 5e^-5 \:I_5
   \end{bmatrix}
 \]
 
  where $\ I_7 $\ is a 7x7 identity matrix, $\ I_{12} $\ is a 12x12 identity matrix, and $\ I_5 $\ is a 5x5 identity
 matrix. The three different identity matrices are the size of the head, flow and leakage
 rate states.

\section{SIMULATION RESULTS}

The state space representation of the leak detection model was developed in Matlab environment. The results obtained are as discussed in this section.

\subsection{Pressure transients due to valve closure}
\begin{figure}[h]
  	  \centering
  		\includegraphics[width=0.5\textwidth]{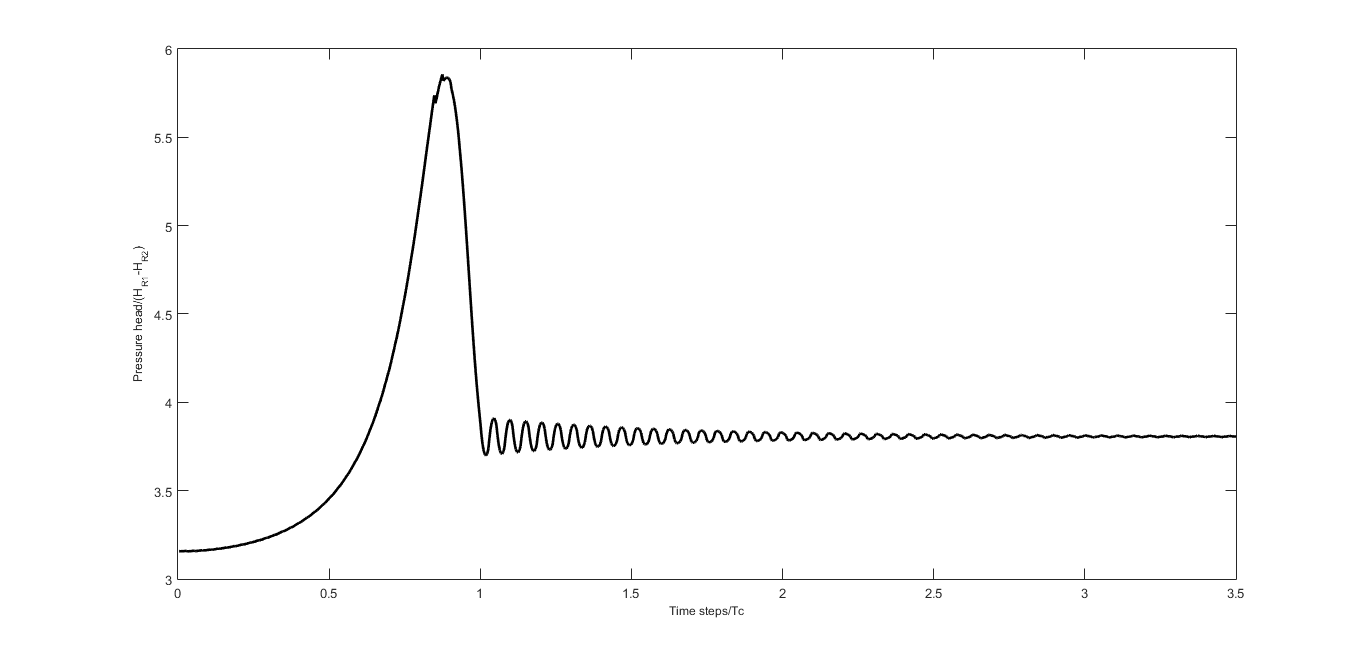}
  	  \caption{Fluid transients developed through valve closure, with no burst-leakage}
  	  \label{figure:f80}
      \end{figure}

 It can be observed [figure \ref{figure:f80}] that the closure of valve results an increase in pressure from the static steady state pressure head of 30$\ (m) $\ to a maximum of 60$\ ( m) $\ (due to water hammer or compressibility effects). The peak pressure corresponds to near about full valve closure time of 20s. The pressure wave there after attenuates to a mean value of 40$\ ( m) $\ due to fluid friction and energy dissipation in pipe material. However, the energy lost as heat dissipation is not modeled. We then compared our results with the model of Lesyshen \cite{11} for validation.

\begin{figure}[h]
  	  \centering
  		\includegraphics[width=0.5\textwidth]{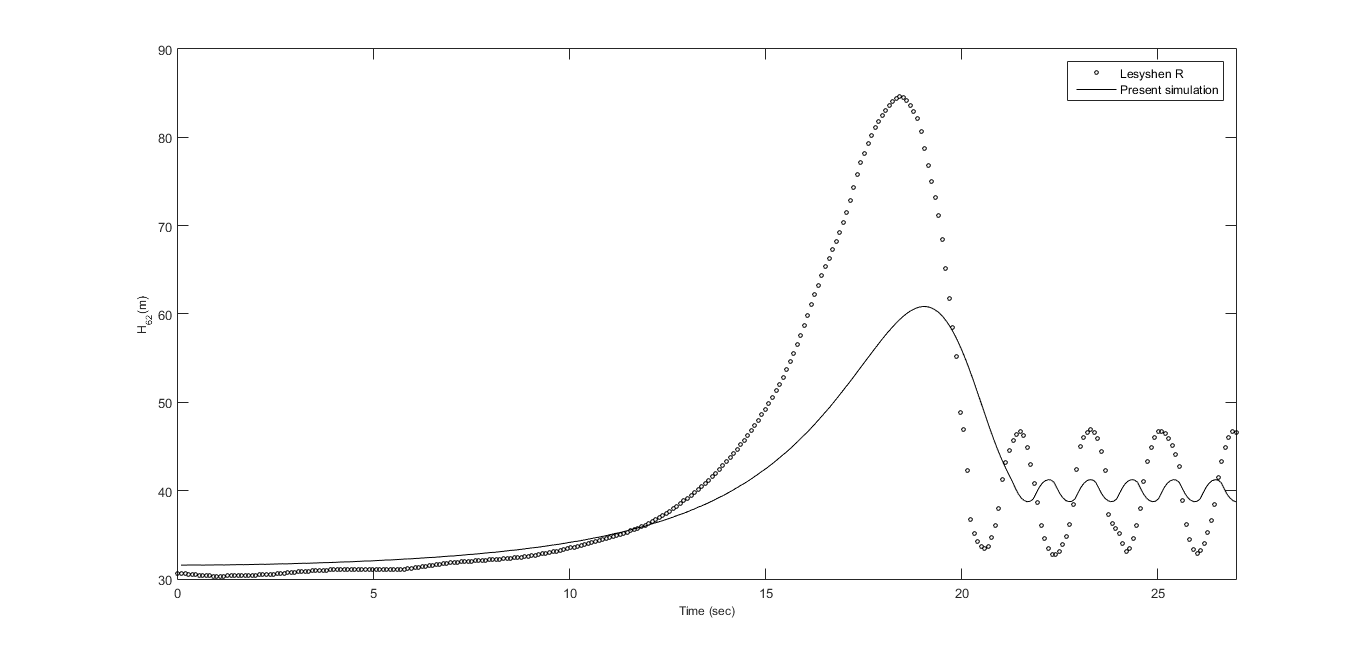}
  	  \caption{Fluid transients developed through valve closure, with
  	  no burst-leakage: validation}
  	  \label{figure:f9}
      \end{figure}

 \par Difference in peak pressure recorded can be attributed to the nature of modeling for transient coefficient of discharge $\ (C_{d}) $\ . We have considered the variation in hydraulic diameter in the calculation for $\ C_{d} $\ , [Equation \ref{equation:aa36}]. Moreover, Lesyshen \cite{11} has accepted that his model over predicts peak pressure based on rigid water column theory.
\par
\begin{figure}[h]
  	  \centering
  		\includegraphics[width=0.5\textwidth]{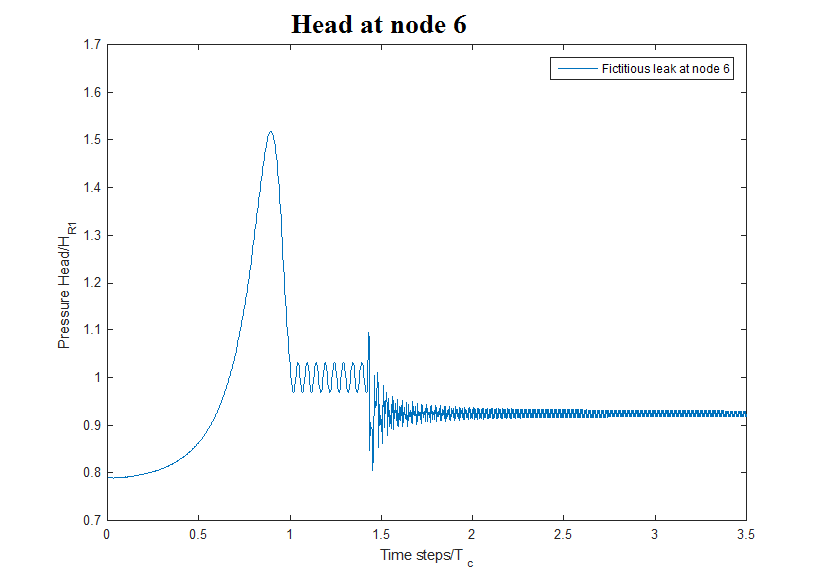}
  	  \caption{Simulation of controlled fictitious burst-leakage at  node 6}
  	  \label{figure:ff9}
      \end{figure}

 Later we introduced a fictitious leak at a specified node (node 6) at the 400 th timestep to employ the additional characteristic equations discussed in section 5.5., to completely introduce all the set of equations of leakage in to the model. The attenuated wave further develops secondary transients on the event of burst leakage (400 th timestep). The results of this fictitious burst leakage is shown in figure \ref{figure:ff9} ( Head at node 6 represented in terms of non-dimensionalized pressure values against non-dimensionalized time). It is clear that this secondary transients tends to die out faster due to the pressure being released after a burst leakage.\\

\par A  fully deterministic model applied among all the eligible nodes for burst.  Later the results were compared with a simple probabilistic model of all the nodes eligible for burst. The instantaneous pipeline burst is characterized by sudden variation in the recorded pressure values to valve closure time (both non-dimensionalized) , as clearly evident in results.
\par
\begin{figure} 
    \centering
  \subfloat[Head at node 2]{%
       \includegraphics[width=0.5\linewidth]{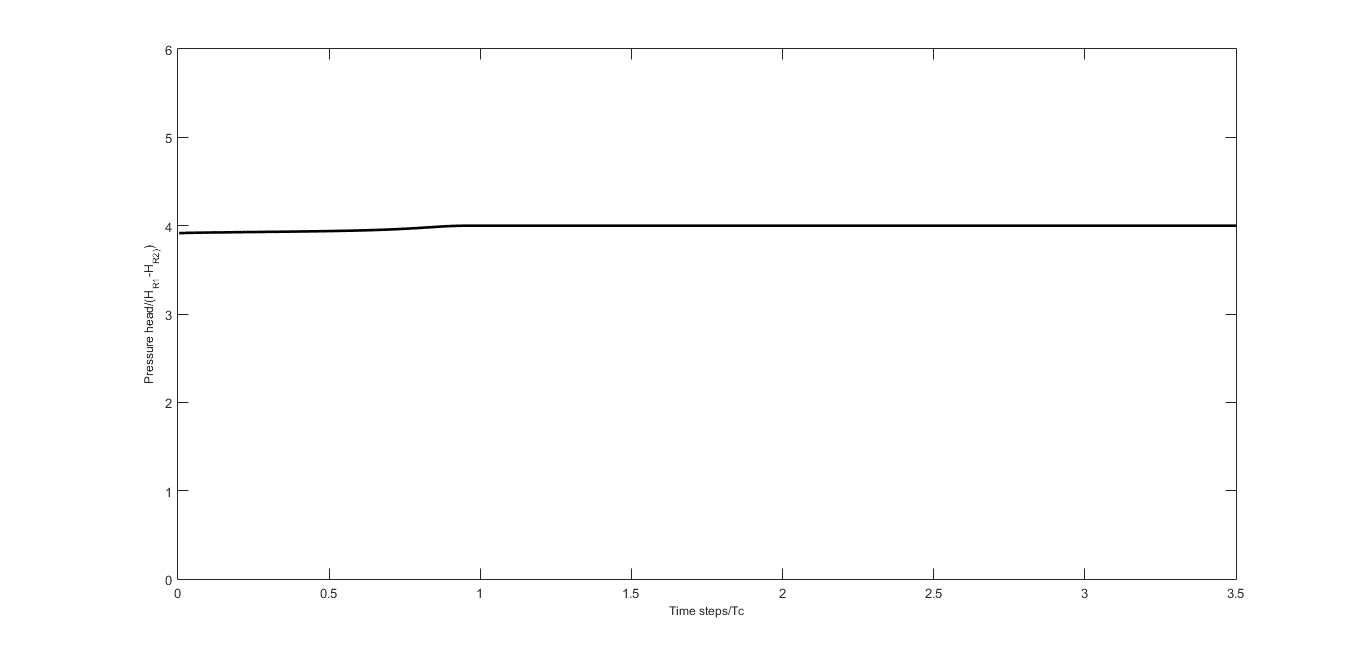}}
    \label{1a}\hfill
  \subfloat[Head at node 3]{%
        \includegraphics[width=0.5\linewidth]{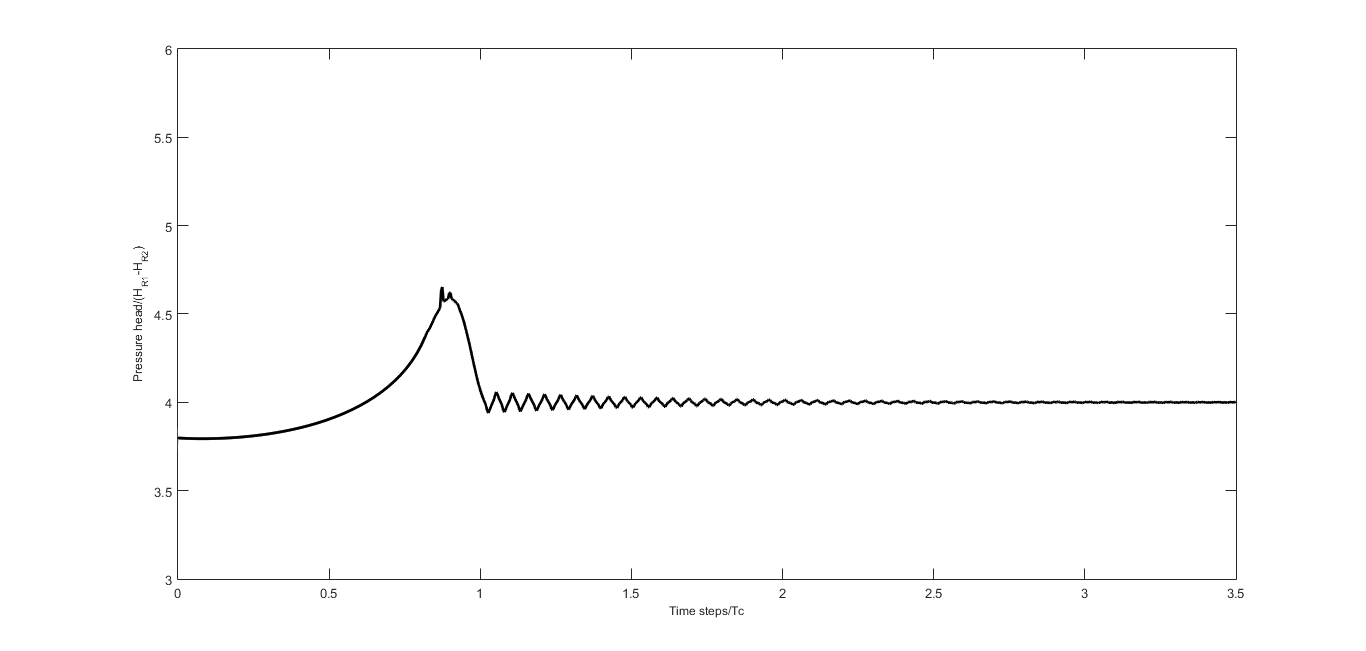}}
    \label{1b}\\
  \subfloat[Head at node 4]{%
        \includegraphics[width=0.5\linewidth]{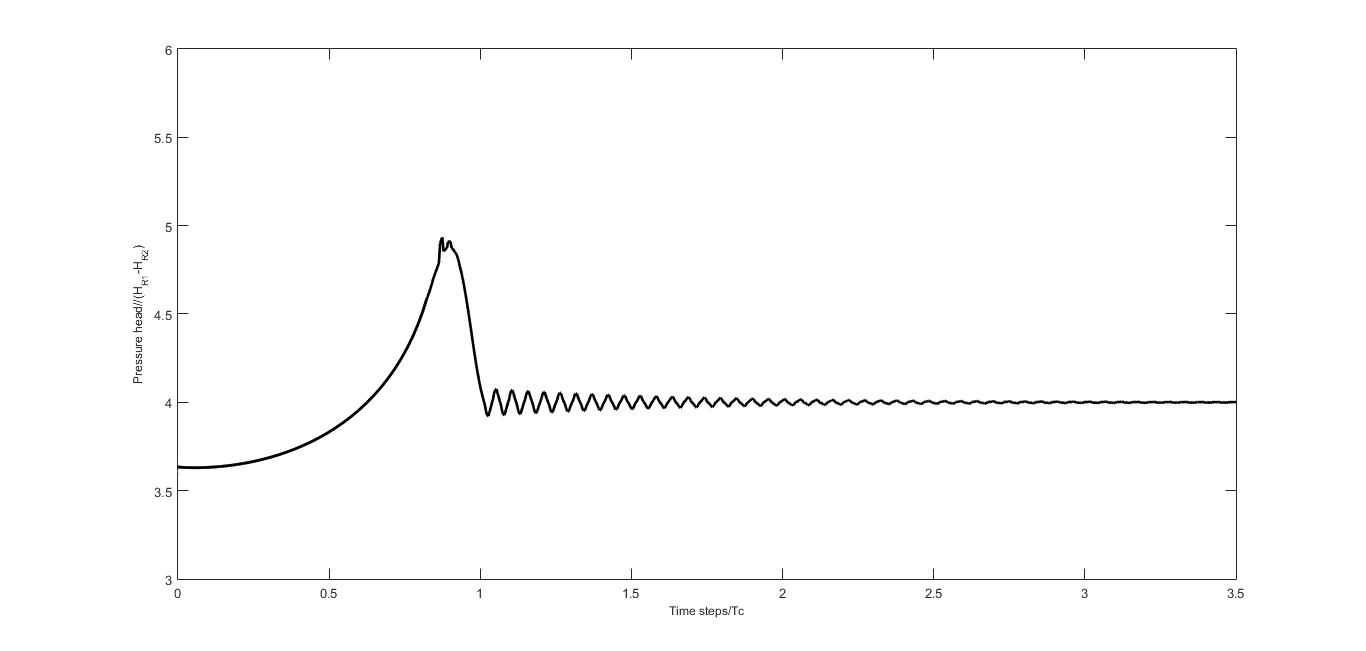}}
    \label{1c}\hfill
  \subfloat[Head at node 5]{%
        \includegraphics[width=0.5\linewidth]{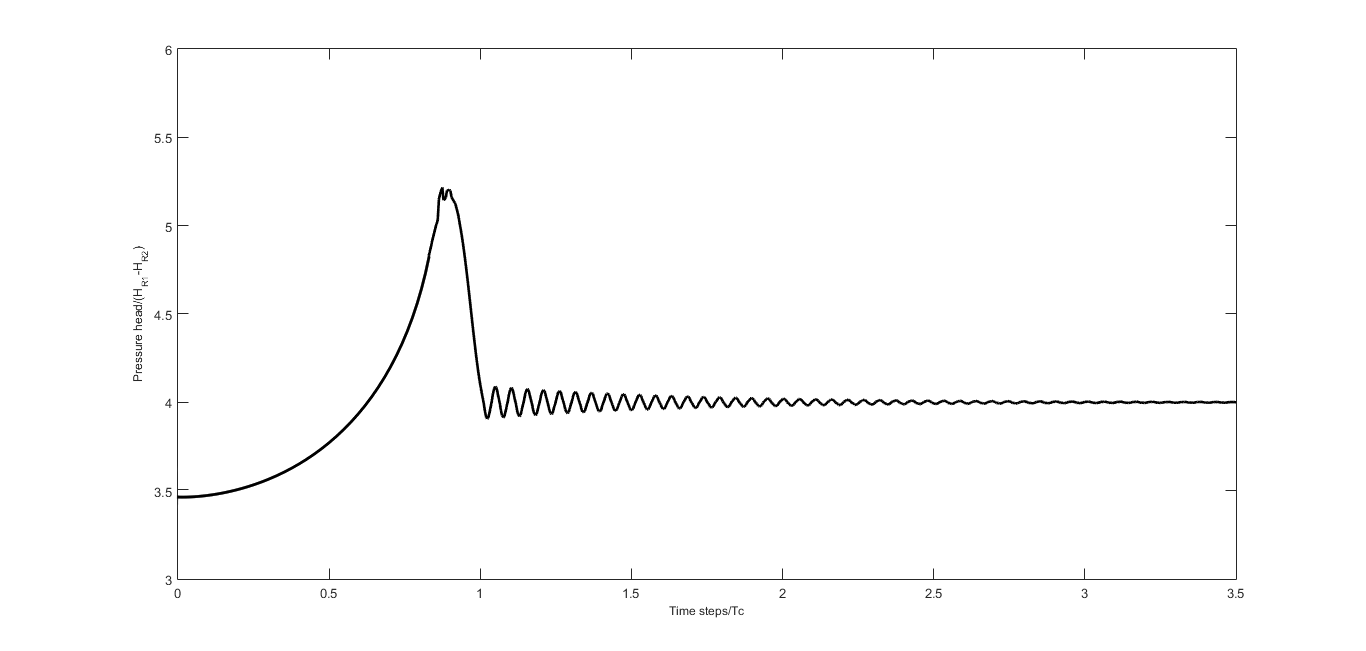}}
     \label{1d} \\
      \subfloat[Head at node 6]{%
             \includegraphics[width=0.5\linewidth]{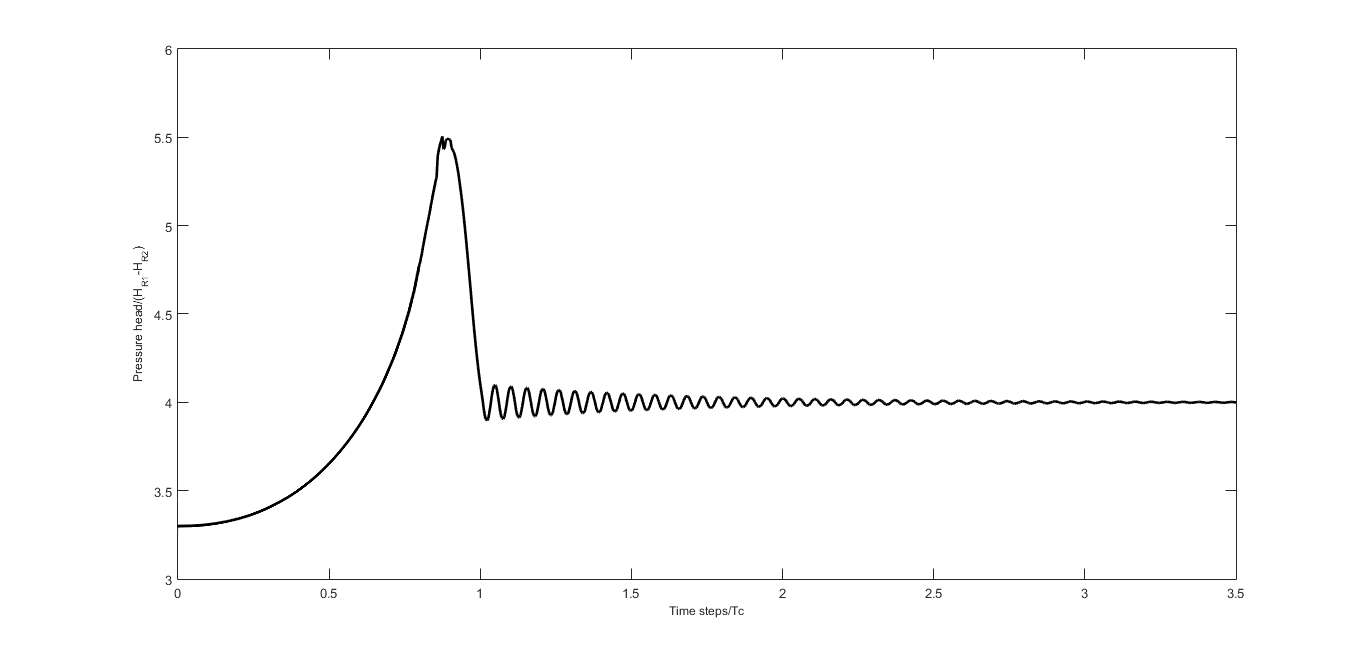}}
         \label{1e}\hfill
        
  \caption{Pressure transients with out burst recorded at internal nodes} 
  \label{fig:fig} 
\end{figure}

 The pressure heads in transient flow simulations in all the internal nodes recorded over time is as shown in figure \ref{fig:fig}. Here the highest peak pressure are recorded at the downstream nodes since they are closer to the valve.
\par Among these nodes, the plausible breakage nodes were those which recorded pressure heads greater than 80\% yield stress in the deterministic model. Meanwhile, we allowed for failure to only those nodes with the high probability. It was assigned based on simple random coin tossing probability. 

\begin{figure} 
    \centering
  \subfloat[Head at node 4]{%
       \includegraphics[width=0.6\linewidth]{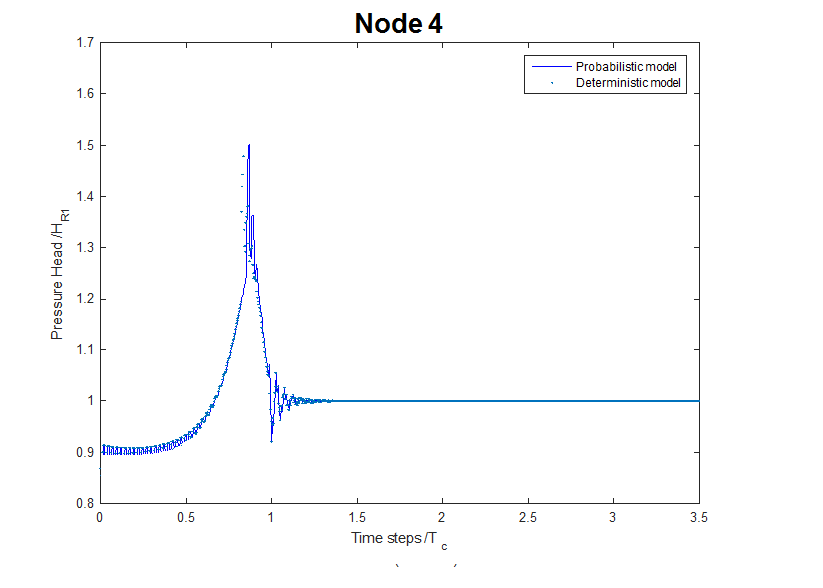}}
    \label{1a}\hfill
  \subfloat[Head at node 5]{%
        \includegraphics[width=0.6\linewidth]{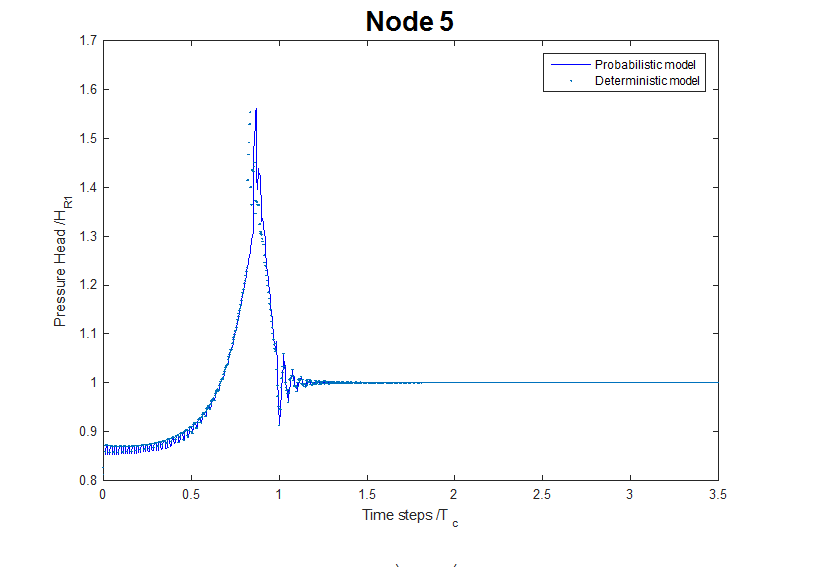}}
    \label{1b}\\
  \subfloat[Head at node 6]{%
        \includegraphics[width=0.6\linewidth]{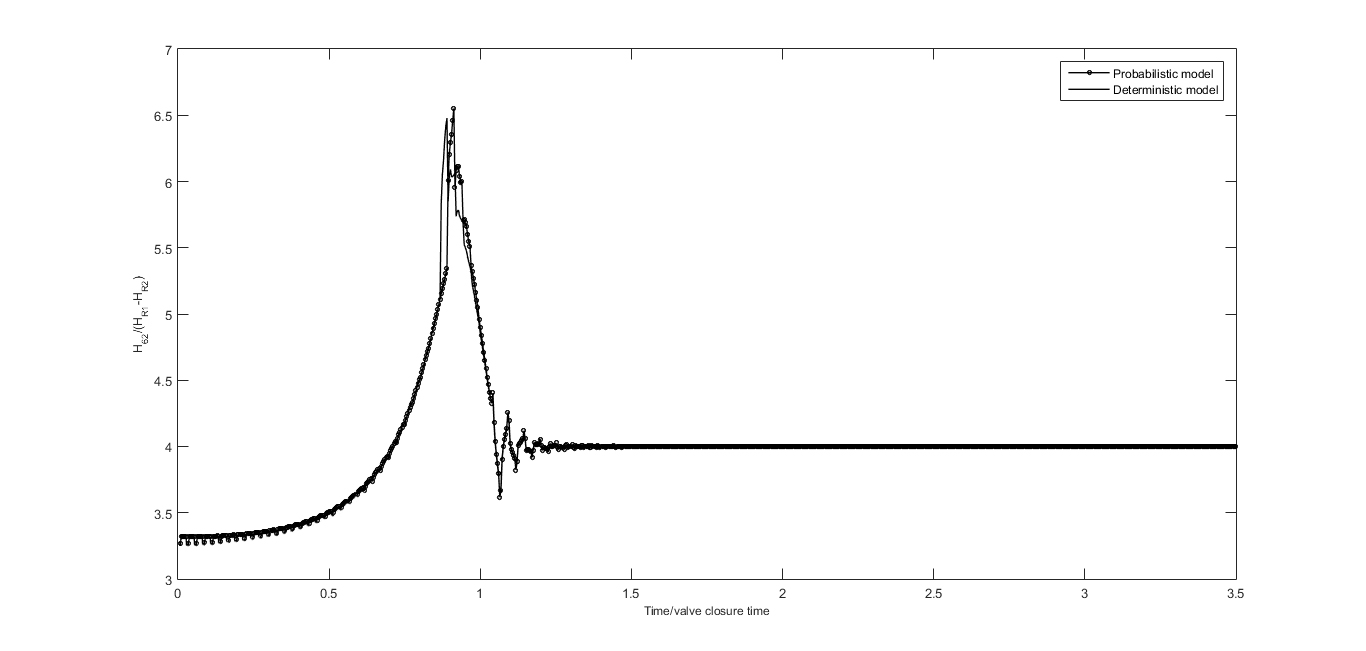}}
    \label{1c}

 \caption{Comparison of transients predicted by the two different models} 
  \label{fig:fig1} 
\end{figure}

A comparison of the transients predicted by the two models are given in figure \ref{fig:fig1}. It is clear from the above figures that the two models are in close agrement both quantitatively and qualitatively in modeling of burst leakage and transients created. However, the probabilistic model saves a node (node 3) among the set of plausible nodes from burst . In all the above results, we observed that the burst to happen close to the full valve closure. 

Figure below shows the actual amount of leak rate in the interior nodes. It is clear that nodes 4,5 and 6 burst and the actual amount of leakage calculated using either models with constant leak area was asymptotically 0.0063 [$\ m^3/sec $\ ] for each leaking node.

\begin{figure} 
    \centering
  \subfloat[Leak rate at node 2]{%
       \includegraphics[width=0.5\linewidth]{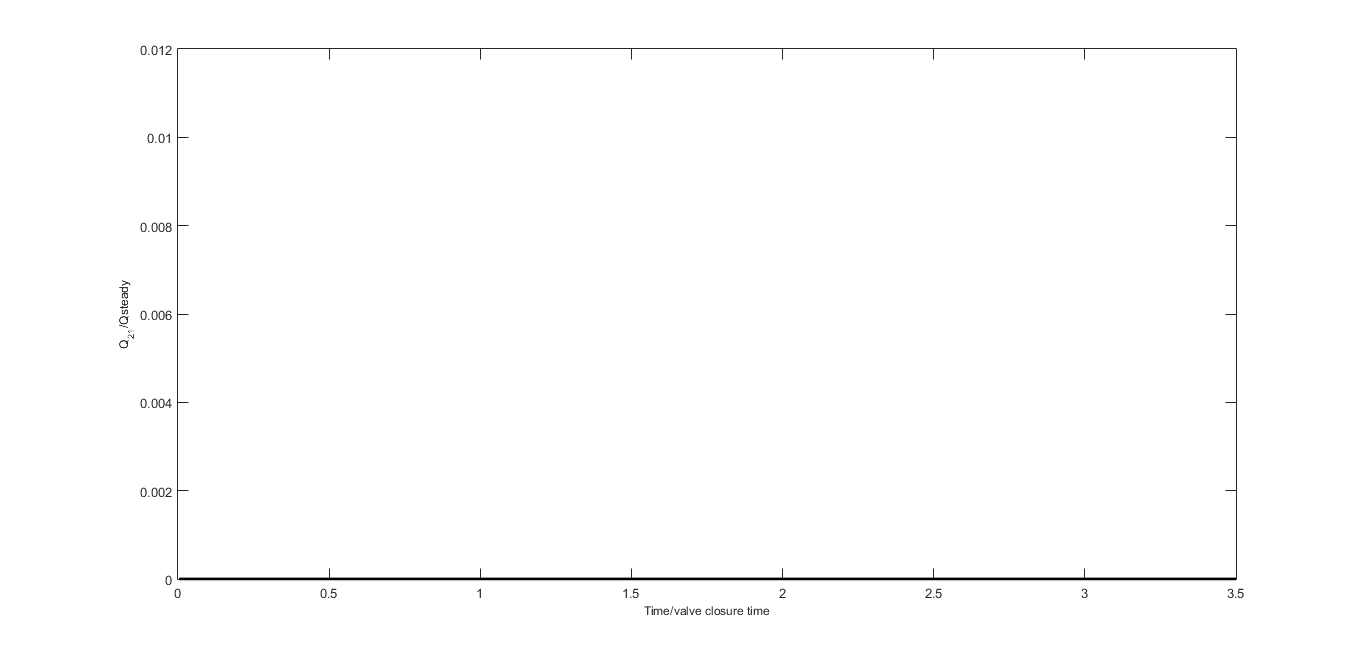}}
    \label{1a}\hfill
  \subfloat[Leak rate at node 3]{%
        \includegraphics[width=0.5\linewidth]{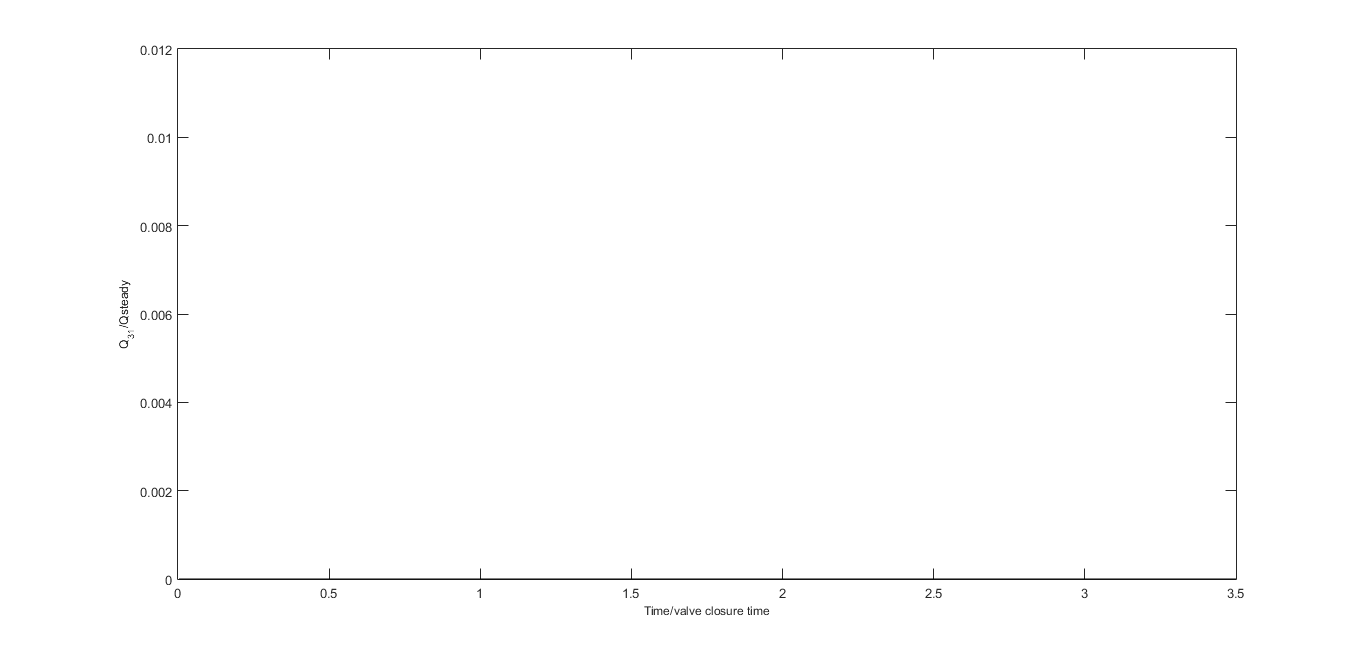}}
    \label{1b}\\
  \subfloat[Leak rate at node 4]{%
        \includegraphics[width=0.5\linewidth]{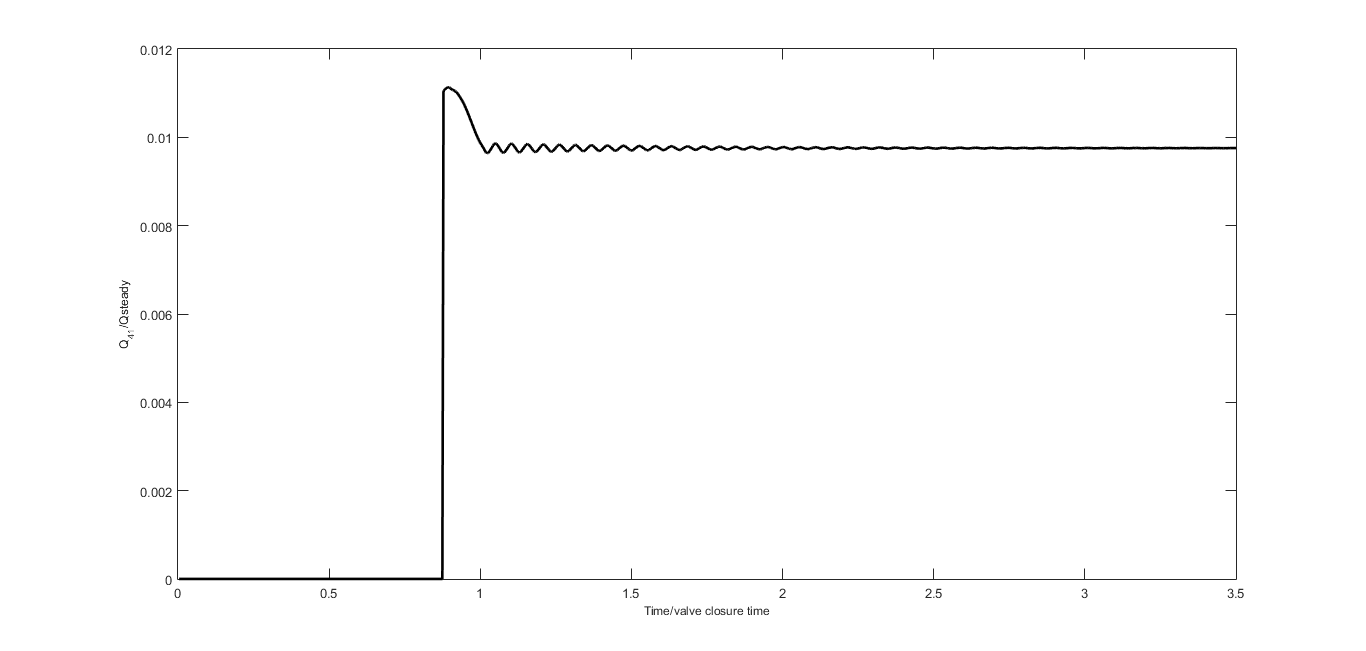}}
    \label{1c}\hfill
  \subfloat[Leak rate at node 5]{%
        \includegraphics[width=0.5\linewidth]{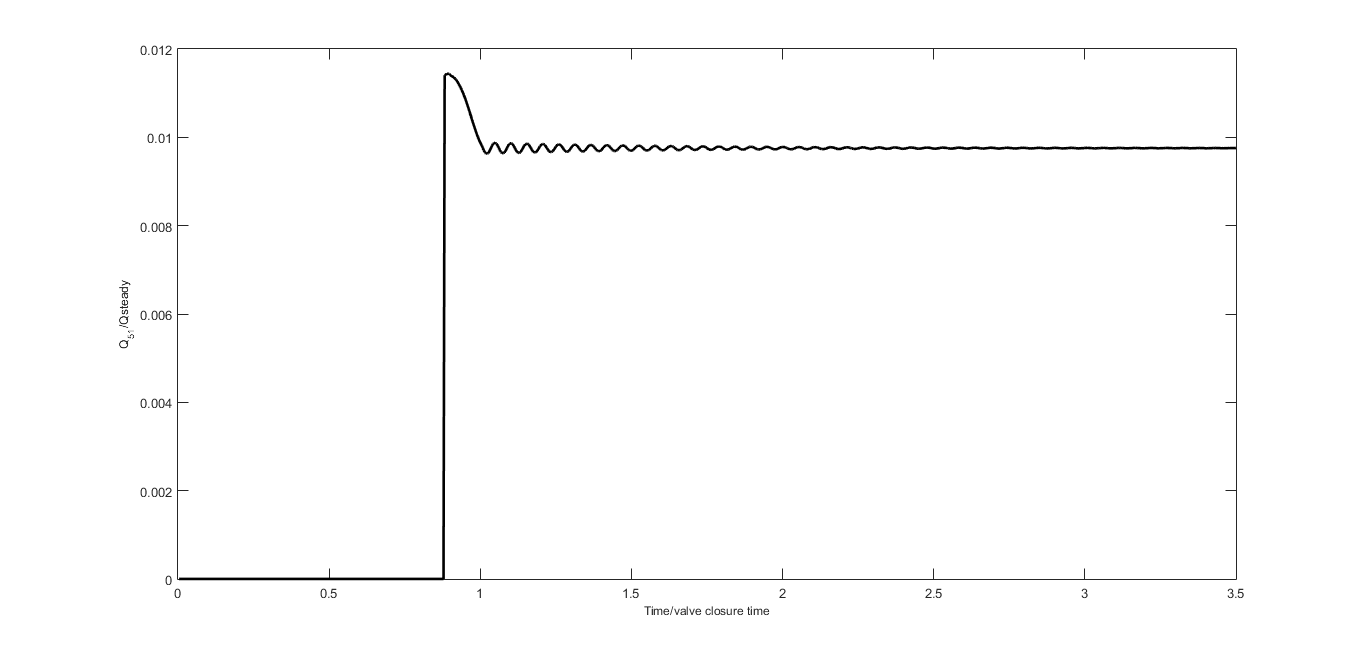}}
     \label{1d} \\
      \subfloat[Leak rate at node 6]{%
             \includegraphics[width=0.5\linewidth]{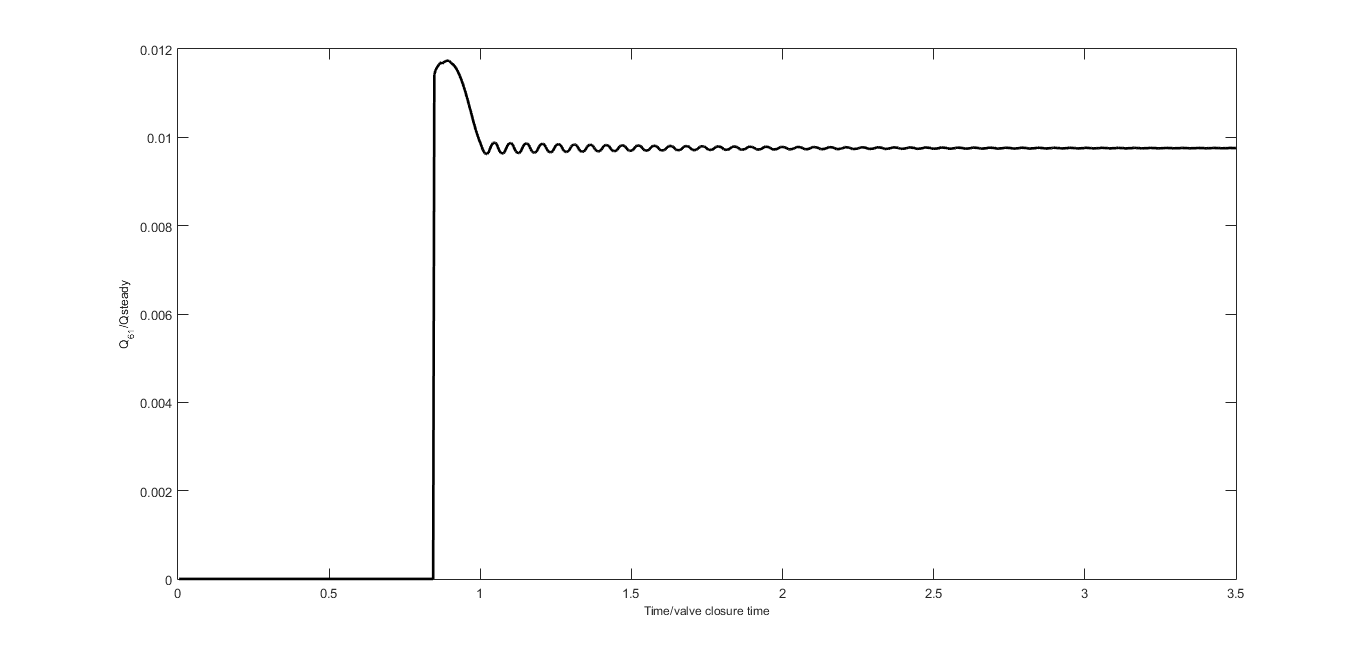}}
         \label{1e}
       
  \caption{Leak rate at interior nodes with hydraulic modeling} 
  \label{fig:fig2} 
\end{figure}

\subsection{Leak prediction using Extended Kalman Filter approach}  

In the EKF approach, pressure head values calculated at nodes 1 and 7 were combined with white noise (with variance .04) to simulate the measurements. And it is assumed that there is zero leakage at the boundary nodes. The results shows that the pressure heads at nodes 4, 5 and 6 rises and attains a peak value due to valve closure and then the transients tend to die out faster and reached a steady state value. 
\par
\begin{figure} 
    \centering
  \subfloat[Leakage rate at node 2]{%
       \includegraphics[width=0.5\linewidth]{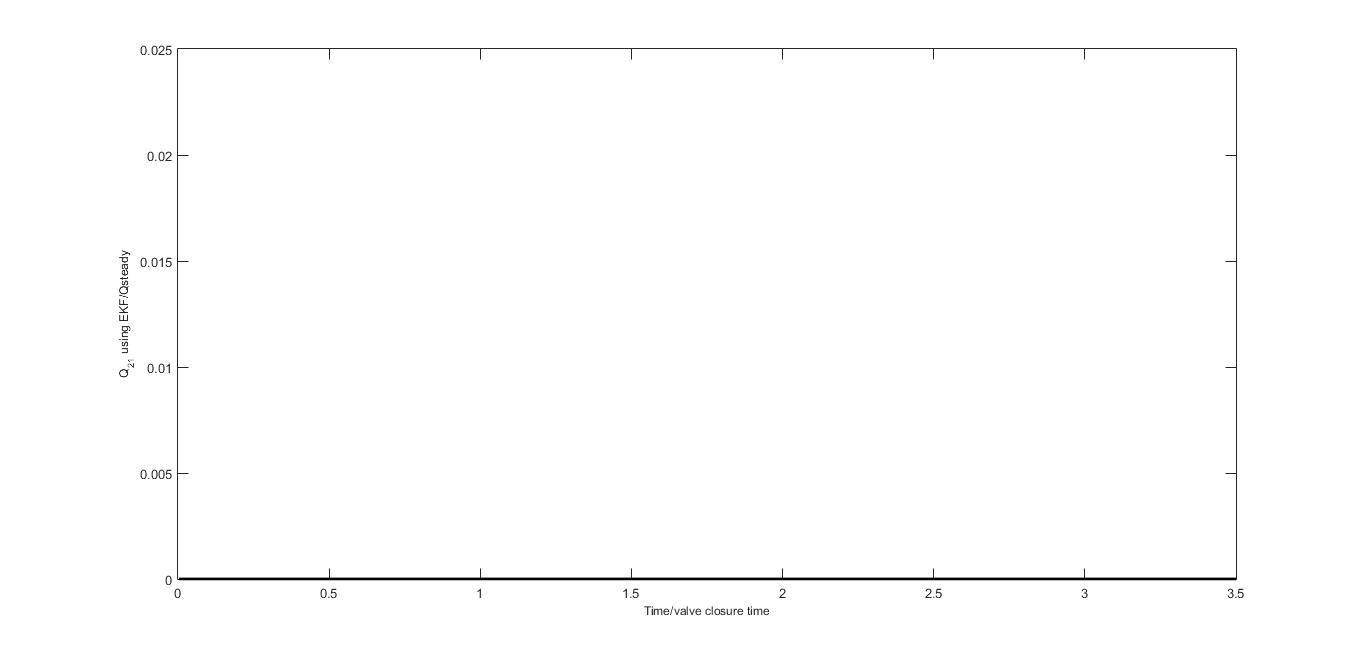}}
    \label{1a}\hfill
  \subfloat[Leakage rate at node 3]{%
        \includegraphics[width=0.5\linewidth]{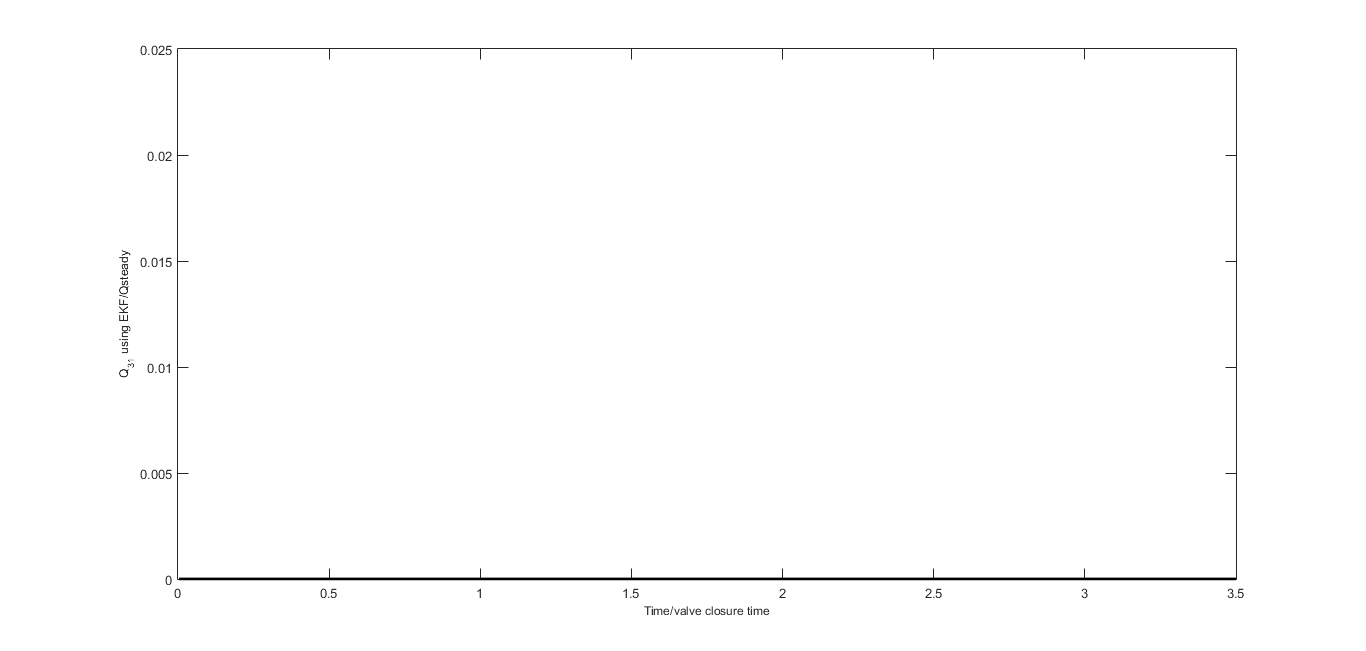}}
    \label{1b}\\
  \subfloat[Leakage rate at node 4]{%
        \includegraphics[width=0.5\linewidth]{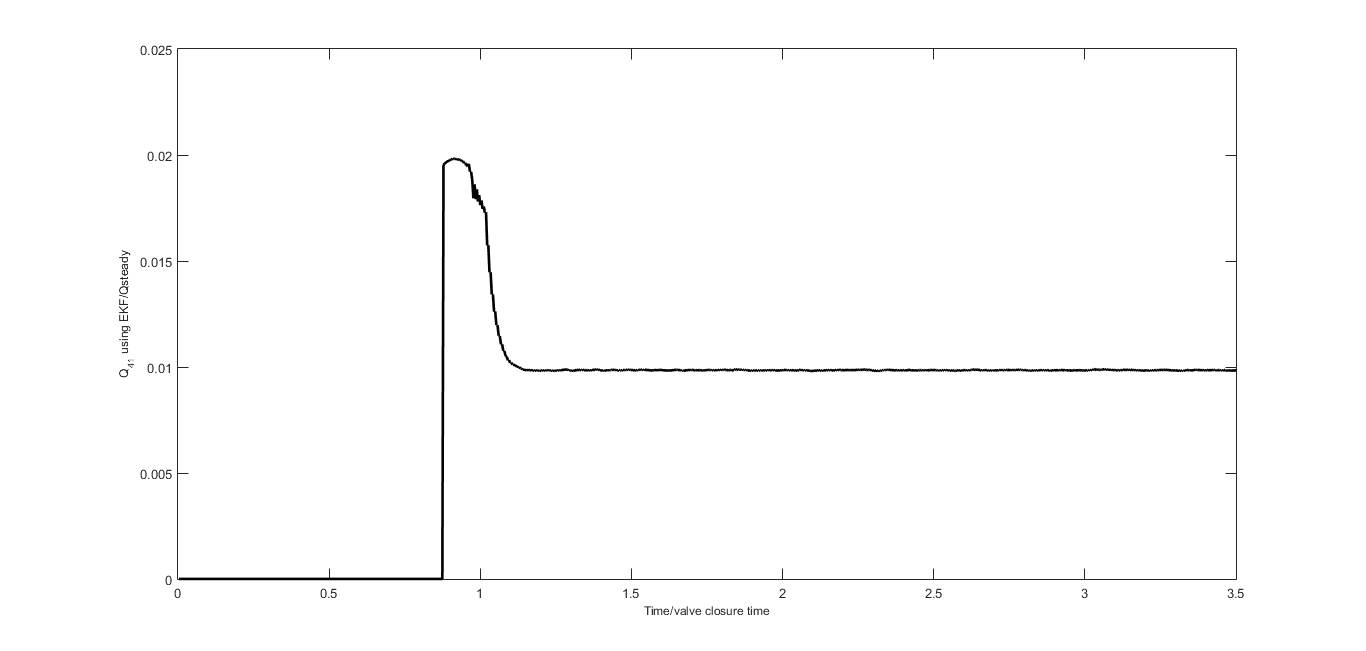}}
    \label{1c}\hfill
  \subfloat[Leakage rate at node 5]{%
        \includegraphics[width=0.5\linewidth]{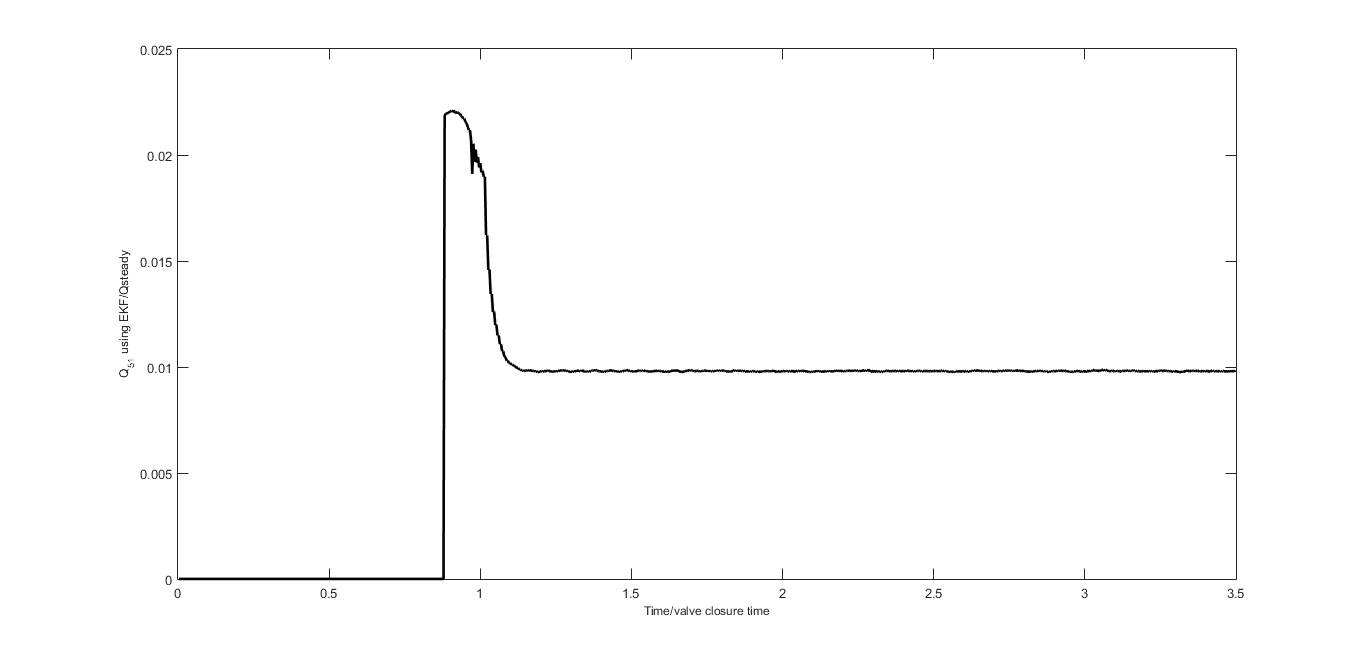}}
     \label{1d} \\
      \subfloat[Leakage rate at node 6]{%
             \includegraphics[width=0.5\linewidth]{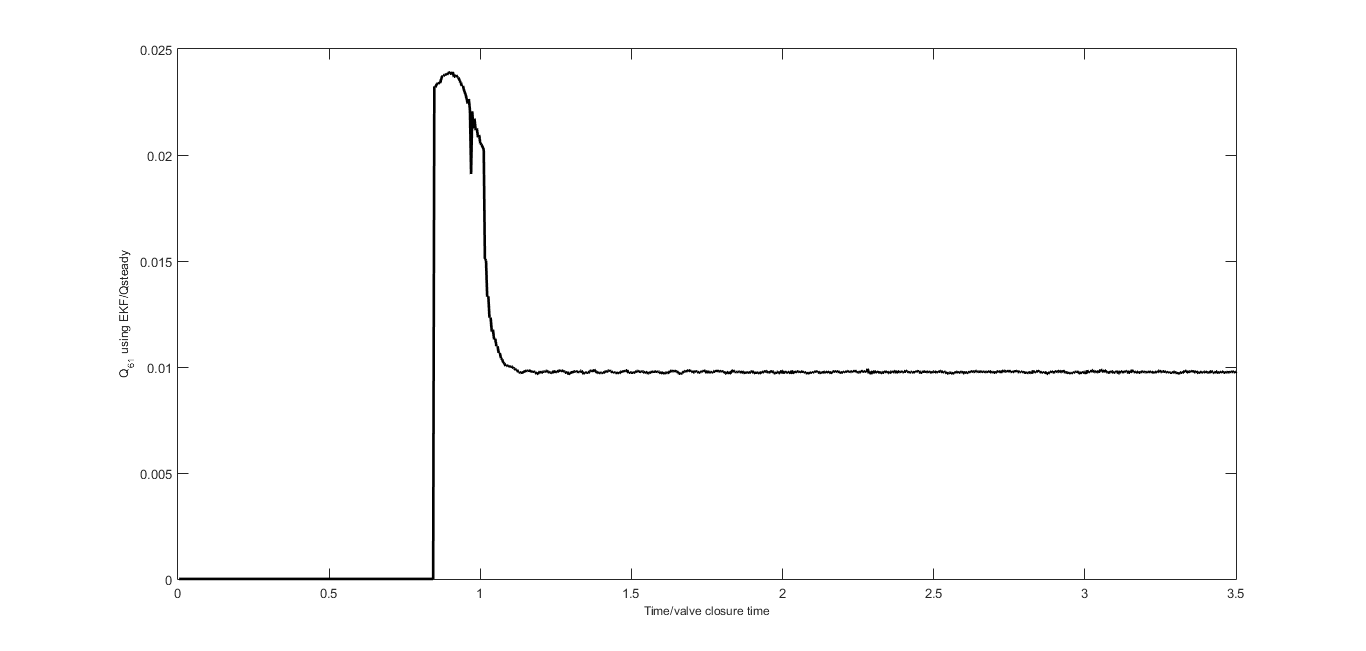}}
         \label{1e}
  
  \caption{Prediction of Leakage rate at interior nodes on applying EKF} 
  \label{fig:fig4} 
\end{figure}

 Using EKF in prediction, the estimated leakages were found at node 4, 5 and 6. Figure. \ref{fig:fig4},  shows the predicted leakage for all of time steps at nodes 2, 3, 4, 5 and 6.

The real leakage was positioned at nodes 4, 5 and 6 and the leak rate was aymptotically  0.0062 [$\ m^3/sec $\ ] at node 4 , 0.0063 [$\ m^3/sec $\ ] at node 5 and 0.00635 [$\ m^3/sec $\ ] at node 6. The position of leakage predicted by system identification in Kalman filter was also at nodes 4, 5 and 6 and the predicted value of 
leak rate over time at node 4 was 0.00622 [$\ m^3/sec $\ ] , at node 5 was 0.0063 [$\ m^3/sec $\ ] and at node 6 was 0.006358 [$\ m^3/sec $\ ] asymptotically. It was found that the predicted  leak rate closely follows the true leak rates obtained from the fluid dynamic model. Kalman prediction of leak involved system identification of the leak nodes as well as estimates of leak rates. At nodes 2 through 6, if the estimated hoop stress was greater than 80 \% of yield stress, we assumed the nodes to burst.\\          

\section{CONCLUSION}
\normalsize A model for burst leakage in pipeline was developed based on Method of Characteristics (MOC). MOC was found to be robust for different input criteria we employed for the simulations. On validating our initial results with Lesyshen \cite{11}, we found that the transients developed were greatly influenced by the nature of equation chosen for coefficient of discharge $\ (C_d)$\ for flow through controlled restrictions such as valves in the pipeline.
Allowing only selected nodes to leak based on a simple probabilistic method, we found it performed only marginally different from a purely deterministic model.
\par On burst, the models predicted secondary transients which dies out rapidly due to immediate release in pressure. For both the probabilistic and deterministic models, the transients due to burst were observed close to complete closure time of the valve.
\par For the failure nodes, the burst leak rate settled around a constant steady state value of 0.0063 [$\ m^3/s $\ ] (20 s), after an immediate momentary hike in leak rate. This tendency is also observable in the numerical scheme with EKF. This steady state leak rate is the solution for the present case, since we have not allowed for the crack area to propagate. If one has to find the actual leak rate under creep for the pipe material, additional equations for the variation of $\ \lambda $\ (leak area propagation) should also be included in the model.
It may be concluded that the methodology described was successful in identifying leakage.\\

\section*{ACKNOWLEDGEMENT}

The authors gratefully acknowledge the technical support in problem definition by Product Engineering Services, Wipro Pvt. Ltd., Kochi, Kerala.

\end{document}